\documentclass[preprint,12pt]{elsarticle}

\usepackage[english]{babel}

\usepackage[letterpaper,top=2cm,bottom=2cm,left=3cm,right=3cm,marginparwidth=1.75cm]{geometry}

\usepackage{graphicx}
\usepackage{xcolor}
\usepackage{upgreek}
\usepackage[colorlinks=true, allcolors=blue]{hyperref}
\usepackage{comment}
\usepackage{booktabs}
\usepackage{array}

\definecolor{jmlBlue}{RGB}{0,163,224}
\newcommand{\bl}[1]{\textcolor{blue}{#1}}
\usepackage{amssymb}
\usepackage{amsmath}
\begin{document}

\begin{frontmatter}



\title{Hybrid classical-quantum communication networks}


\author{Joseph M. Lukens$^{a,b,c}$, Nicholas A. Peters$^b$, Bing Qi$^{d,e}$\corref{cor1}} %% Author name

\affiliation[1]{organization={Elmore Family School of Electrical and Computer Engineering and Purdue Quantum Science and Engineering Institute, Purdue University},%Department and Organization
            city={West Lafayette},
            state={Indiana},
            postcode={47907}, 
            country={USA}}

\affiliation[2]{organization={Quantum Information Science Section, Computational Sciences and Engineering Division, Oak Ridge National Laboratory},%Department and Organization
            city={Oak Ridge},
            state={Tennessee},
            postcode={37831}, 
            country={USA}}

\affiliation[3]{organization={Research Technology Office and Quantum Collaborative, Arizona State University},%Department and Organization
            city={Tempe},
            state={Arizona},
            postcode={85287}, 
            country={USA}}
            
 \affiliation[4]{organization={New York University Shanghai, NYU-ECNU Institute of Physics at NYU Shanghai},%Department and Organization
            addressline={567 West Yangsi Road}, 
            city={Shanghai},
            postcode={200126}, 
            country={China}}           

 \affiliation[5]{organization={State Key Laboratory of Precision Spectroscopy, School of Physical and Material Sciences,
East China Normal University},%Department and Organization
            city={Shanghai},
            postcode={200062}, 
            country={China}}   

\cortext[cor1]{Corresponding author.
E-mail address: bing.qi@nyu.edu\\
This manuscript has been co-authored by UT-Battelle, LLC, under contract DE-AC05-00OR22725 with the US Department of Energy (DOE). The US government
retains and the publisher, by accepting the article for publication, acknowledges that the US government retains a nonexclusive, paid-up, irrevocable, worldwide
license to publish or reproduce the published form of this manuscript, or allow others to do so, for US government purposes. DOE will provide public access to
these results of federally sponsored research in accordance with the DOE Public Access Plan (http://energy.gov/downloads/doe-public-access-plan).}

%% Abstract
\begin{abstract}
Over the past several decades, the proliferation of global classical communication networks has transformed various facets of human society. Concurrently, quantum networking has emerged as a dynamic field of research, driven by its potential applications in distributed quantum computing, quantum sensor networks, and secure communications. This prompts a fundamental question: rather than constructing quantum networks from scratch, can we harness the widely available classical fiber-optic infrastructure to establish hybrid quantum-classical networks? This paper aims to provide a comprehensive review of ongoing research endeavors aimed at integrating quantum communication protocols, such as quantum key distribution, into existing lightwave networks. This approach offers the substantial advantage of reducing implementation costs by allowing classical and quantum communication protocols to share optical fibers, communication hardware, and other network control resources—arguably the most pragmatic solution in the near term. In the long run, classical communication will also reap the rewards of innovative quantum communication technologies, such as quantum memories and repeaters. Accordingly, our vision for the future of the Internet is that of heterogeneous communication networks thoughtfully designed for the seamless support of both classical and quantum communications.
\end{abstract}

%% Keywords

\end{frontmatter}



\section{Introduction}

The prediction by Charles Kao and George Hockham in 1966 that ultralow-loss silica fiber could enable long-distance information transmission has had a profound impact on society~\cite{Kao1966Dielectric}. During the past five decades, sustained efforts by scientists and engineers around the world have transformed fiber-optic communication from simple point-to-point links that span tens of kilometers~\cite{Schwartz1978Atlanta, Berry1978Optical} into an extensive, interconnected global network that now supports most Internet traffic.

More recently, the concept of a quantum Internet has emerged through theoretical proposals and preliminary experiments~\cite{Kimble2008Quantum, Wehner2018Quantum}. The ultimate goal is to connect remote quantum information processing devices by exchanging both quantum and classical information. Harnessing the power of quantum superposition and entanglement, the quantum Internet could enable groundbreaking applications, including distributed quantum computing, quantum-secure communication, and distributed quantum sensing. However, quantum systems are notoriously prone to decoherence due to interactions with the environment. Currently, photons in the optical domain, where blackbody radiation is negligible, are considered to be the most promising information carriers for long-distance quantum communication. Consequently, the future quantum Internet is most likely to be built upon fiber-optic infrastructures, along with free-space optical links. Thus, it is essential to understand to what extent we can leverage existing classical fiber-optic communication technologies and infrastructure to implement various quantum communication protocols.

The primary objectives of a communication system are to transmit information efficiently, reliably, and securely. Technological advances in this field are largely driven by efforts to address two key challenges: extending communication distances and optimizing network resource allocation among multiple users. The evolution of classical fiber-optic communication networks can be categorized into three main eras \cite{agrell2016roadmap, Winzer2018Fiber}. (1) \textit{Regenerated Direct-Detection Era}. This phase extended communication distances using classical regenerators. During this initial period, communication capacity was primarily limited by the bandwidth of transceivers. (2) \textit{Amplified Dispersion-Managed Era}. Characterized by the widespread adoption of wavelength-division multiplexing (WDM) and practical optical amplifiers, such as erbium-doped fiber amplifiers (EDFAs), this era introduced more cost-effective and scalable solutions. (3) \textit{Amplified Coherent Communication Era}. Marked by the introduction of coherent optical receivers, this phase enabled longer regenerator spacing and significantly improved spectral efficiency. Looking ahead, while continuing to increase the capacity of single-fiber links may not fully meet the exponential demand for data transmission, parallel spatial paths through space-division multiplexing (SDM)~\cite{li2014space} present a promising solution.

Quantum communication, in contrast, remains in its infancy. Researchers are actively exploring the theoretical foundations of a quantum Internet, addressing aspects ranging from the physical layer to network control architecture. However, experimental advances have lagged considerably behind theoretical progress. Even implementing the most basic communication task---transmitting a qubit over a lossy channel deterministically and with high fidelity~\cite{hermans2022qubit}---poses immense challenges, let alone developing more advanced protocols. To date, the primary exceptions are protocols compatible with probabilistic quantum communication. A prominent example is quantum key distribution (QKD)~\cite{bennett2014quantum, ekert1991quantum}, which allows remote users to establish a shared cryptographic key using a probabilistic quantum channel in conjunction with an authenticated classical channel. In fact, QKD and related quantum cryptographic protocols are expected to be among the most impactful and practical applications of quantum networks in the near term.

On the one hand, given the limited near-term applications of quantum networking, a viable path toward commercialization involves the integration of quantum communication protocols into existing fiber-optic networks. Significant research has been dedicated to this approach, from investigating the coexistence of classical and quantum signals within the same fiber via WDM technology to exploring systems that enable simultaneous classical and quantum communications. On the other hand, innovative technologies developed for a future quantum Internet—such as low-loss fiber-optic devices, optical memory, highly sensitive detectors, \bl{and distributed entanglement}—could also bring transformative advancements to classical communication networks. Our vision for the Internet’s future is that of a heterogeneous communication network, thoughtfully designed to seamlessly support both classical and quantum communications.

This paper provides a comprehensive review of ongoing research efforts focused on integrating quantum communication protocols, such as QKD, into existing fiber-optic communication networks. Section \ref{sec:fundamentals} presents a concise overview of the fundamental concepts in both classical and quantum communication. Section \ref{sec:integration} addresses the integration of quantum communication protocols into conventional fiber-optic networks. Section \ref{sec:outlook} concludes with a brief summary.

\section{Fundamentals of classical and quantum networking}
\label{sec:fundamentals}
Two main challenges in large-scale communication networks are ensuring reliable information transmission over lossy and noisy channels, and optimizing the efficient use of network resources. In this section, we present concise overviews of key technologies that have been developed and proposed in both classical fiber-optic and emerging quantum networks.

\subsection{Fundamentals of classical fiber-optic networks}

Given that approximately ninety percent of the information received by the human brain is visual, the history of optical communication can, in a broad sense, be traced back to the origins of human civilization. Purpose-built optical communication systems emerged at least two thousand years ago, allowing information to be transmitted over long distances using smoke signals, beacon fires, flags, and other methods. For example, as early as 440 BC, the Peloponnesians used lighthouses to alert Athens of enemy attacks~\cite{Thucydides2019History}. Similarly, in ancient China, beacon towers along the Great Wall served a comparable purpose~\cite{Chen2024Discovery}.

\begin{figure}[t]
	\includegraphics[width=.8\textwidth]{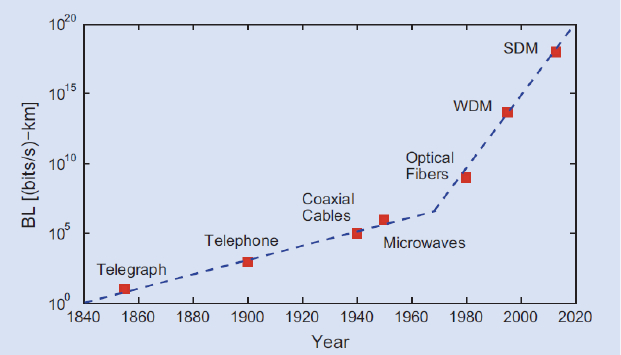}
    \centering
    %%	\captionsetup{justification=raggedright, singlelinecheck=false }
	\caption{Information capacity of communication networks measured in the bit rate--distance product (BL) has grown exponentially during the period of 1840--2015. (Image reproduced with permission from a Creative Commons Attribution 4.0 International License (\url{http://creativecommons.org/licenses/by/4.0/}) \cite{agrawal2016optical}.)} 
	\label{fig:1}
\end{figure}

A major advantage of optical communication is its vast bandwidth capacity. However, early forms of long-distance optical communication were limited in application due to several factors: communication rates were constrained by slow encoding processes, and reliability was compromised by unpredictable weather conditions. Widespread long-distance optical communication only became feasible after Corning scientists Robert Maurer, Peter Schultz, and Donald Keck developed ultralow-loss silica fiber in 1970, following predictions by Charles Kao and George Hockham in 1966~\cite{Kao1966Dielectric}. Since then, the information capacity of fiber-optic networks---measured by the bit rate--distance product (BL)---has grown exponentially, as illustrated in Fig.~\ref{fig:1} (note the acceleration in growth rate around 1977)~\cite{agrawal2016optical}. In the following section, we introduce the fundamentals of classical fiber-optic communication networks.

\subsubsection{The basic structure of an optical communication system}

A communication system comprises three major components: the transmitter, receiver, and communication channel. This paper focuses specifically on optical communications using standard single-mode fibers.

\begin{figure}[t]
	\includegraphics[width=.8\textwidth]{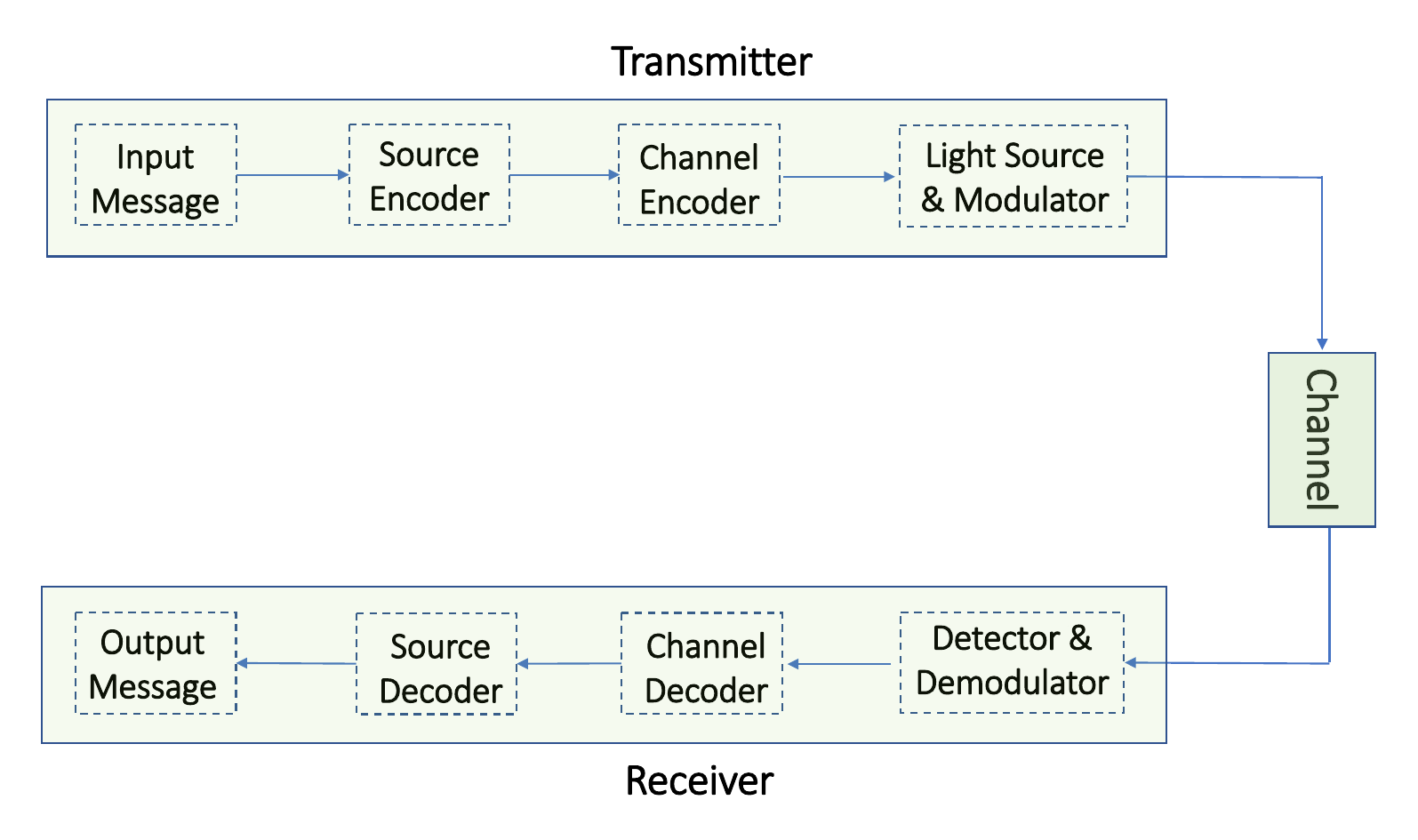}
    \centering
    %%	\captionsetup{justification=raggedright, singlelinecheck=false }
	\caption{The basic structure of an optical communication system.} 
	\label{fig:2}
\end{figure}

Figure~\ref{fig:2} illustrates the basic structure of a typical optical communication system. On the transmitter side, a source encoder can be employed to compress the input message, enhancing communication efficiency. According to Shannon's source coding theorem~\cite{shannon1948mathematical}, in the asymptotic case, an $N$-bit message from an independent and identically distributed (i.i.d.) source can be compressed into approximately $NH(X)$ bits with minimal risk of information loss, where $H(X)$ is the source entropy. On the receiver side, a source decoder is used to reconstruct the original message from the compressed data.

For improved system reliability, a channel encoder can be used at the transmitter, with a corresponding channel decoder at the receiver. This is particularly valuable in long-distance communications, where the channel may be lossy and noisy. Channel coding introduces redundancy, enabling error correction that supports reliable data recovery even under poor signal-to-noise ratio (SNR) conditions.

After channel encoding, data in the electrical domain is transferred to the optical domain by modulating specific properties of a light source, such as intensity, phase, polarization, spatial mode, or a combination of these. The modulation bandwidth, or the spectral width of the signal, is directly proportional to the data transmission rate.

In early fiber-optic communication systems, broadband sources such as light-emitting diodes (LEDs) were commonly used due to their simple driver requirements and low cost~\cite{lee1980recent}. However, to make more efficient use of the optical bandwidth, in modern coherent WDM systems, narrowband sources with a linewidth smaller than the modulation bandwidth are typically employed~\cite{winzer2012high}. This approach maximizes data-carrying capacity and improves overall system performance. Despite this evolution, LEDs remain relevant for short-distance links and are anticipated to play a key role in emerging technologies such as the Internet of Things (IoT) and 6G wireless networks~\cite{ren2021emerging}.

\begin{figure}[t]
	\includegraphics[width=.8\textwidth]{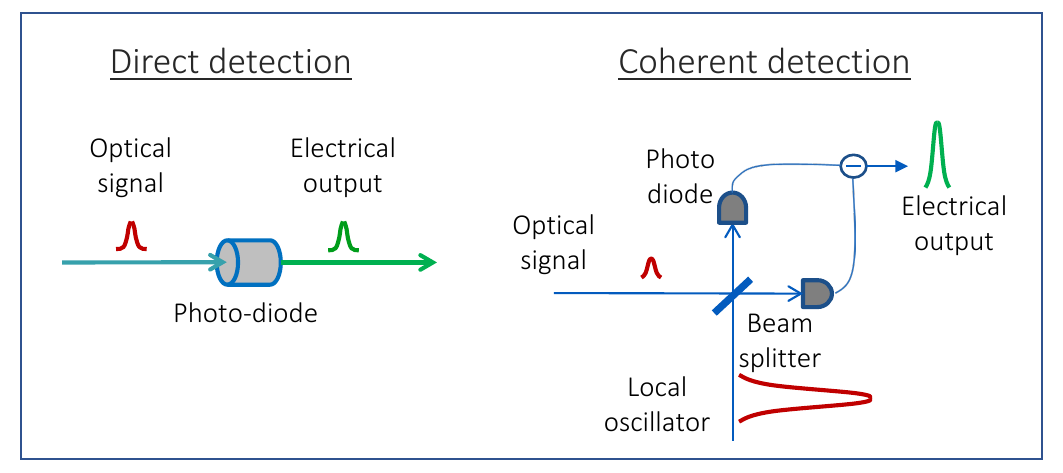}
    \centering
   	\caption{Direct direction and coherent detection in optical communications.} 
	\label{fig:3}
\end{figure}

The role of the optical detector is to convert the optical signal received from the channel back into an electrical form. Two commonly used detection schemes are direct detection and optical coherent detection, as shown in Fig.~\ref{fig:3}. In direct detection, a photodetector measures the intensity of the received optical signal, but the phase information is lost. In contrast, in the coherent detection scheme, the received optical signal beats with the local oscillator (LO), to create an interference signal that is then converted into an electrical signal through photodetection. Both the intensity and phase information can be recovered in coherent detection. 

Direct detection is simpler in structure and easier to implement, making it advantageous in certain applications. However, optical coherent detection is becoming the dominant choice in long-distance fiber-optic communications due to its ability to extend regenerator spacing and enhance spectral efficiency \cite{kikuchi2015fundamentals}. A key advantage of coherent detection is the intrinsic filtering provided by the LO: only signals that match the LO in polarization, spectral, and spatiotemporal mode contribute significantly to the output electrical signal. This filtering capability is especially valuable in quantum communication over dense WDM fiber networks carrying strong classical traffic \cite{qi2010feasibility,kumar2015coexistence,eriksson2019wavelength}, as well as in free-space optical links \cite{heim2014atmospheric}. In those cases, noise photons in other modes can be effectively suppressed in coherent detection.

\subsubsection{Fiber loss management}

An effective communication channel ensures error-free transmission of messages from the sender to the receiver. In fiber-optic communication, fiber loss is a key factor influencing the range of transmission. The advent of low-loss optical fibers transformed fiber-optic networks, establishing them as the backbone of the modern Internet. In recognition of this milestone, Charles Kuen Kao was awarded half of the Nobel Prize in Physics in 2009 ``for groundbreaking achievements concerning the transmission of light in fibers for optical communication''~\cite{Kao2009}. As shown in Fig. \ref{fig:1}, the rate of exponential growth in communication capacity significantly accelerated around the 1970s with the global deployment of fiber-optic networks~\cite{agrawal2016optical}.

Generally, the average power of an optical signal traveling through a fiber link decreases exponentially with the fiber length $L$ as described by $P_{\text{out}} = P_{\text{in}} \times 10^{-\alpha L/10}$, where $P_{\text{in}}$ and $P_{\text{out}}$ are the input and output power, and $\alpha$ is the attenuation coefficient of the optical fiber, measured in dB/km. For a detector with a given noise equivalent power (NEP), the SNR decreases exponentially with increasing fiber length, thereby restricting the maximum communication distance. Since the 1970s, significant efforts have focused on reducing fiber loss. Today, commercial silica-based optical fibers can achieve attenuation below 0.160 dB/km at 1550 nm \cite{ten2016ultra}, with the lowest demonstrated attenuation of 0.1419 dB/km at 1560 nm, primarily limited by Rayleigh scattering at this wavelength \cite{tamura2018first}. Further reductions in fiber loss are possible with innovative designs such as hollow-core fiber \cite{poletti2014nested}.

Even with ultralow-loss fibers, such as those with an attenuation of 0.14 dB/km, communication over long fiber links becomes impractical over distances of several hundred kilometers. For instance, in a 500 km link, the attenuation reaches approximately 70 dB, meaning that, on average, only one photon out of every ten million would successfully pass through the fiber. To overcome this limitation and extend the reach of optical communications, regenerators and optical amplifiers have been introduced.

An ideal regenerator restores a degraded signal by performing three critical functions: reamplification, reshaping, and retiming (3R) \cite{horvath2020optical}. The most common type of regenerator employs optical-electrical-optical (OEO) conversion, compensating for fiber loss and correcting signal distortion by converting the optical signal to an electrical one, processing it, and then regenerating the optical signal. OEO regenerators were initially adopted in span-by-span regenerated fiber-optic communication systems, where the channel capacity was limited by the interface rates of the OEO converter \cite{Winzer2018Fiber}. However, this approach presents significant challenges in modern WDM networks due to its inability to adapt to different modulation formats and the requirement for a dedicated regenerator for each wavelength channel. Optical amplifiers offer an alternative approach to loss management by amplifying the entire WDM band in the optical domain, thereby rendering the communication channel optically ``transparent'' \cite{agrawal2012fiber}.

Although the concept of optical amplification dates back to the 1960s, its widespread application in fiber-optic communication networks did not begin until the 1990s, following the development of EDFAs~\cite{mears1987low, desurvire1987high}. Since then, the EDFA has become the most widely deployed optical amplifier in commercial systems. Doped with the rare-earth element erbium, standard EDFAs offer low-noise amplification in the 1530--1565 nm wavelength range, which corresponds to the low-loss C-band of silica-based optical fibers.

To accommodate growing bandwidth demands, the operational range of EDFAs was later extended to the long-wavelength L-band (1565--1625 nm). The resulting C+L-band configuration significantly increases the number of WDM channels that can be supported a single fiber strand~\cite{cantono2020opportunities}. While efforts to extend amplification beyond the C+L-band have yet to see widespread deployment, such advancements are expected to play a significant role in future capacity upgrades of optical networks~\cite{hoshida2022ultrawideband}.

Typically, commercial EDFAs offer gains in the range of 20–30 dB, enabling signal transmission over distances up to approximately 100 km in standard single-mode fiber without regeneration.  However, state-of-the-art EDFAs can achieve gains exceeding 40 dB, as demonstrated by commercially available systems~\cite{Thorlabs}.

Optical amplifiers can boost signal power through stimulated emission, but they also introduce additional noise through spontaneous emission. Amplified spontaneous emission (ASE) is a primary factor contributing to SNR degradation in fiber-optic systems. The Noise Figure (NF), defined as the ratio of input SNR to output SNR in dB, quantifies the noise introduced by optical amplifiers. According to quantum mechanics \cite{caves1982quantum}, the theoretical minimum NF for any phase-insensitive amplifier, such as an EDFA, is 3 dB, with typical commercial EDFAs ranging between 5 and 7 dB at low input power. This fundamental noise imposes a limit on the number of cascaded optical amplifier stages before regeneration becomes necessary. 

\subsubsection{Fiber dispersion management}

Once the distance limitation caused by fiber attenuation was overcome through the introduction of optical amplifiers, chromatic dispersion emerged as one of the most significant impairments in fiber-optic communication systems \cite{charlet2016ultra}.

As a light pulse propagates through an optical fiber, chromatic dispersion causes different frequency components of the pulse to travel at slightly different speeds. This results in pulse broadening, which leads to intersymbol interference (ISI) as adjacent pulses begin to overlap. ISI imposes a limit on the maximum error-free communication distance. The challenge becomes more pronounced at higher transmission rates: as the transmission rate increases, pulse duration shortens, and spectral width broadens, causing the dispersion-limited distance to decrease quadratically with the symbol rate. For instance, the chromatic dispersion of standard single-mode fiber (SMF) is approximately 17 ps/nm/km at 1550 nm. When the transmission rate increases from 2.5 Gb/s to 10 Gb/s, the dispersion-limited transmission distance decreases from around 800 km to just 60 km, even with chirp-free light sources \cite{Winzer2018Fiber}. This underscores the importance of dispersion management in high-speed communication systems.

To mitigate the effects of chromatic dispersion, dispersion-shifted fiber (DSF) was introduced in the 1980s. DSF is designed to have near-zero chromatic dispersion ($\sim$0~ps/nm/km) at 1550 nm. However, in WDM networks, near-zero dispersion proved problematic. Optical signals from different wavelength channels could interact through nonlinear processes such as four-wave mixing (FWM), generating new wavelengths. These nonlinear interactions are more pronounced in fibers with low dispersion, particularly with evenly spaced wavelength channels. The photons generated through FWM often fall within the signal bands of the original wavelengths, potentially causing coherent interference and degrading system performance.

To address these issues, modern systems use nonzero dispersion fibers (NZDF) with carefully managed dispersion characteristics. In NZDF-based systems, fibers with opposite signs of chromatic dispersion are cascaded to balance overall dispersion across the link. This approach minimizes pulse broadening while maintaining enough ``local'' dispersion along the fiber to suppress nonlinear effects like FWM \cite{Winzer2018Fiber}. Such dispersion management strategies have been critical in enabling high-capacity and long-distance optical communication networks.

\subsubsection{Multiuser lightwave networks}
\label{sec:multiuser}
As a communications medium, optical fiber can support multiple degrees of freedom (DoFs)---spatial, spectral, temporal, and polarization---for low-loss transmission, each of which can be decomposed into modes that are uniquely assigned to pairs of users communicating over shared cables. Such modal-based multiplexing is possible even in the SMF common in long-distance communications, for the ``single-mode'' descriptor refers to limitation of transverse spatial modes only. Indeed, by eliminating the water peak absorption around 1383~nm, modern SMF expands the already broadband support of legacy SMF to a remarkable range; specifically, G.652.D class optical fibers---as standardized by the International Telecommunciation Union (ITU)~\cite{ITU2016}---achieve attenuation $\leq$0.4~dB/km over the entire 1310--1625~nm wavelength band, or more than 44~THz of usable bandwidth. With such broadband support in a single fiber, it is therefore unsurprising that the spectral DoF in the form of WDM continues to represent the leading method for supporting multiple users in lightwave networks.

Ubiquitous in optical networks for several decades now~\cite{Ishio1984, Brackett1990, Bergano1996}, WDM has historically been divided into ``coarse'' WDM (CWDM) and ``dense WDM'' (DWDM). ITU specifies the former as 20~nm-wide channels~\cite{ITU2003} and the latter as channel spacings ranging from 12.5~GHz to multiples of 100~GHz, with center frequencies equal to integer channel multiples away from 193.1~THz ($\sim$1552.5~nm)~\cite{ITU2020}. From a practical perspective, the primary difference between CWDM and DWDM networking centers on cost and performance tradeoffs. The much tighter spectral tolerances of DWDM require lasers and filters with GHz-scale accuracies, but in turn can support many more users: for example, the 12.5~GHz DWDM grid corresponds to more than 900 channels in the C+L-band, compared to a mere five CWDM channels in that same range.

An even more flexible form of DWDM has emerged in recent years---namely ``elastic'' or ``flex-grid'' optical networking~\cite{Gerstel2012, Jinno2017}. Rather than relying on fixed channel center frequencies and widths, flex-grid supports spectral channels of a variety of widths, enabling precise allocation of bandwidth to support disparate user needs; e.g., a high-bandwidth demand can receive a single ultrabroad 1~THz channel, while a low-bandwidth user concurrently obtains a 12.5~GHz one---there is no need to force them onto a fixed grid that may leave critical resources underutilized. ITU recommends central frequencies on a 6.25~GHz grid and slot widths in multiples of 12.5~GHz, but otherwise allows for any set of nonoverlapping channels~\cite{ITU2020}. In practice, flex-grid implementations are enabled by technology such as wavelength-selective switches (WSSs)~that combine Fourier-transform pulse shaping~\cite{Weiner2000, Weiner2011} with sub-spot-size gratings for fully programmable control of the amplitude, phase, and output fiber on a frequency-by-frequency basis (limited only by spectral resolution)~\cite{Roelens2008, Ma2021}. With such technology, it is possible to not only realize specific instantiations of flexible DWDM allocations, but also reconfigure slots on demand as bandwidth requirements evolve.

An additional multiplexing approach available to optical networks, time-division multiplexing (TDM)  assigns users slots in the time domain, typically interleaved such that $N$ users each communicating at rate $R_0$ multiplex into a single line rate of $NR_0$. Also enjoying a longstanding history in lightwave communications~\cite{Tucker1988, Spirit1994, Seo1996}, TDM can be divided into electrical and optical versions, distinguished by whether the multiplexing and de-multiplexing operations are performed in the electronic or photonic domains, respectively. In the former case, multiple data streams are combined digitally and modulated onto a optical field; at the receiver, the data stream is detected at the full line rate and then demultiplexed into constituent tributary channels digitally. In the optical approach, the tributary channels are modulated onto optical fields separately, then combined, transmitted, and demultiplexed optically before optical-electrical (OE) conversion at separate detectors. As an entirely digital technique, electrical TDM requires no modifications to the optical architecture, yet the line rate $NR_0$ is then constrained by the bandwidth of OE conversion (either on the modulation or detection front). By contrast, optical TDM requires OE bandwidths only at the tributary rate $R_0$, thus in principle enabling much higher total line rates. Nevertheless, all-optical TDM still requires resources at the full bandwidth $NR_0$, including short pulses, delay lines, and fast switches. Such technological difficulties combined with the continued increase in digital signal processing (DSP) bandwidths have prevented optical TDM from making a significant commercial impact.

Finally, moving beyond standard SMF to novel fiber types is opening up additional multiuser opportunities in SDM, where users are assigned distinct transverse spatial modes in, e.g., the same core of a multimode fiber (MMF) or separate cores in the common cladding of a multicore fiber~\cite{Richardson2013, li2014space, Puttnam2021}. While the cost of system upgrades makes the level of commercial SDM adoption unclear at the moment, SDM's ability to multiply the transmission capacity of individual cables by potentially orders of magnitude suggests a compelling niche in space-constrained and high-bandwidth links.

\subsubsection{All-optical signal processing}
\label{sec:allOptical}
Optical TDM, SDM, and the various flavors of WDM described in Sec.~\ref{sec:multiuser}  form a subset of a wider corpus of techniques known collectively as all-optical signal processing~\cite{Willner2014}. Broadly defined, all-optical signal processing encompasses any communication operation---e.g., multiplexing, impairment compensation, and logic---that is performed entirely in the optical regime with no conversion either to or from the electronic domain. Significant experimental progress has been made on a variety of photonic capabilities conventionally associated with electronics, including demultiplexing and pulse shaping in more complex multiuser paradigms such as optical code-division multiple-access~\cite{Heritage2007}, optical orthogonal frequency-division multiplexing~\cite{Hillerkuss2011}, and optical orthogonal time-division multiplexing~\cite{Soto2013}; wavelength exchange and multicasting~\cite{Bres2009, Wang2010, Biberman2010, Lu2020a}; tunable delay lines~\cite{Okawachi2005, Sharping2005}; and high-speed logic~\cite{Xu2007, Eggleton2012}.

Much of the motivation for pursuing all-optical signal processing stems from the ultrahigh bandwidth of optical signals, allowing for operations in principle as fast as the bandwidth supported in fiber (i.e., 44~THz in G.652.D class SMF as defined above~\cite{ITU2020}). Moreover, the avoidance of OE conversion can potentially reduce loss, improve energy efficiency, and lower latency. Yet actualizing such advantages can be quite difficult in practice, as the history of optical packet switching highlights. Compared to circuit switching, in which bandwidth is allocated and reserved for the duration of any communication session, packet switching routes data dynamically, in a segment-by-segment basis over shared bandwidth~\cite{Roberts1978}. While circuit switching was successful in traditional telephone networks and remains the method of choice in the optical physical layer (cf. flex-grid bandwidth allocation), packet switching has proven critical to the expansion of the Internet, enabling the support of orders of magnitude more users than circuit switching over the same bandwidth. In lightwave networks, routing decisions are conventionally made in the electronic domain: optical packets are detected, stored, and then retransmitted to their appropriate destination. Research into all-optical alternatives flourished in the 1990s, and many clever solutions based on wavelength converters, switches, and delay lines were developed~\cite{Hunter2000, Xu2001, OMahony2001, Yao2001}. Yet the complexity, high loss, and lack of reliable optical storage have prevented them from infiltrating commercial lightwave networks at any meaningful level~\cite{Ramaswami2006}---a situation persisting to today.

With the exception of WDM systems and the continued trend to supplant electrical wires with optical fiber for data transmission on ever-shortening length scales, such as datacenters and supercomputers~\cite{Taubenblatt2012, Miller2017, Cheng2018}, all-optical signal processing techniques have faced similar challenges toward adoption as optical packet switching. And in fact, in modern lightwave communications~\cite{agrell2016roadmap}---particularly in the explosion of coherent encoding formats~\cite{kikuchi2015fundamentals}---DSP has improved to such an extent that the trend has even gone in the opposite direction, with electronics assuming responsibilities previously the purview of photonics. For example, rather than compensating dispersive spreading with optical components such as dispersion-compensating fibers and chirped fiber Bragg gratings, chromatic dispersion is now routinely compensated digitally after OE conversion. More generally, the expansion of forward-error correction \cite{chang2010} has massively relaxed SNR requirements in optical transmission, shifting much of the burden for error-free communications into the digital domain. Interestingly, the destruction of quantum coherence  inherent in photodetection precludes the application of most of these DSP techniques for quantum signals, inspiring us to speculate that quantum networks might finally provide the application needed to push all-optical processing from the lab to the field.

\subsection{Fundamentals of quantum communications}

\subsubsection{Basic concepts}

The primary objective of an optical communication system is to transmit information reliably, that is, deterministically and with negligible error, through a channel that may be both lossy and noisy. In classical optical communication, information (typically measured in bits) is encoded in distinguishable states that are intrinsically resilient to a certain level of channel degradation. This distinguishability allows for perfect state discrimination, enabling the use of classical repeaters based on a measurement-and-regeneration scheme. Furthermore, encoding classical bits onto bright light pulses minimizes signal degradation during optical amplification.

In contrast, quantum communication encodes information in non-orthogonal quantum states, which cannot be perfectly replicated due to the constraints imposed by the quantum no-cloning theorem~\cite{Wootters1982,dieks1982communication}. This fundamental limitation prohibits the use of classical repeaters and optical amplifiers in the majority of quantum communication protocols.

Classical optical communication protocols are typically classified based on their detection methods: direct detection and coherent detection. In contrast, quantum communication protocols are generally divided into two categories based on the dimensionality of the information carriers: discrete-variable (DV) and continuous-variable (CV) protocols. DV protocols utilize information carriers defined in finite-dimensional spaces, such as the polarization or time bin of single photons, whereas CV protocols rely on information carriers defined in infinite-dimensional spaces, like the quadrature components of an optical mode.

The fundamental unit of quantum information is the qubit (the quantum mechanical counterpart to the classical bit), defined in a two-dimensional complex Hilbert space. Given the computational basis states $\vert 0 \rangle$ and $\vert 1\rangle$, a pure qubit state can be expressed as a superposition of the two basis states, $\vert\psi\rangle = a\vert 0\rangle + b\vert 1\rangle$, where $a$ and $b$ are complex numbers that satisfy the normalization condition $\vert a \vert^2 + \vert b \vert^2 = 1$. The unique characteristics of quantum communication, absent in classical communication, along with its associated challenges, stem from quantum superposition and, in the case of multiple photons, quantum entanglement~\cite{horodecki2009quantum}.

In theory, an infinite amount of classical information can be encoded on a qubit, since the above coefficients $a$ and $b$ are complex numbers. However, the amount of accessible information is constrained by the Holevo bound \cite{holevo1973bounds}, which defines the maximum classical information that can be extracted from a quantum system represented by its density matrix $\rho$. For example, a projective measurement of a qubit yields only a single bit of information about the quantum state.

Quantum communication is still in its early stages, with much of the research centered on a fundamental question: how to reliably transmit a qubit through a lossy channel. To tackle this challenge, various quantum repeater schemes have been proposed, including techniques based on entanglement swapping and quantum teleportation \cite{zukowski1993event, bennett1993teleporting}, and quantum error correction codes \cite{fowler2010surface}. These approaches will be discussed in the following sections. Table \ref{tab:comparison} provides a summary of the key differences and similarities between classical and quantum optical communication systems.

%Table one--comparision of classical optical communication and quantum communication
\begin{table}[ht]
\centering
\footnotesize
\caption{Comparison of classical optical communication and quantum communication}
\label{tab:comparison}
\begin{tabular}{@{}|p{2.5cm}|p{5cm}|p{5cm}|@{}}
\hline
Aspect & Classical Communication & Quantum Communication
\\
\hline
Communication Channel & 
Any optically transparent medium, including optical fiber and free space links. & 
Any optically transparent medium, including optical fiber and free space links. \\
\hline
Light Source & Laser; incoherent light source. & 
Laser; incoherent light source; single-photon source; squeezed state of light; entangled light.
 \\
\hline
Information Encoding & 
Actively encodes classical information on orthogonal states of light. &  Actively encodes classical or quantum information on orthogonal or non-orthogonal states of light;
passively generates information during measurement process (e.g., entanglement-based QKD). \\
\hline
Photon Detection & 
Phase-insensitive intensity detection; phase-sensitive coherent detection. & 
Phase-insensitive single-photon detection; phase-sensitive coherent detection. \\
\hline
Channel Loss Management & 
Bright light pulses; optical amplification; classical regenerators & Quantum repeaters (under development).
\\
\hline
\end{tabular}
\end{table}

\subsubsection{Quantum repeaters}

Ultimately, it is desirable to achieve fault-tolerant communications of quantum information at a global scale.  Eventually such a system could correct both loss and operational errors, such that the quantum communications system has performance that is comparable and compatible with the applications that utilize quantum and conventional network resources. In that respect, the quantum communications part of the network would run along with the conventional data communications in the background enabling applications that are only possible due to the two paradigms working in concert.  The development of the quantum repeater and how to best integrate it into existing infrastructure is a topic of accelerating research interest.

The idea of that something other than phase-insensitive amplification is required for large-scale high-performance quantum communications is a direct consequence of the no-cloning theorem~\cite{Wootters1982,dieks1982communication}. However, the need developed historically in the context of a concrete use case.  That use case was the protocol of QKD, which is covered subsequently, where a great deal of work went into developing better protocols and systems to transmit higher rates of keys over increasingly larger-loss channels.  A recurring output was that the rate decreased exponentially with the increasing channel length.  It was some time before it was established that a quantum repeater, or something similar, was rigorously needed to extend the performance of quantum key exchange, even with prefect hardware due to a fundamental rate-loss tradeoff \cite{takeoka2014fundamental}.  Later, this result was generalized more broadly for quantum communications, further cementing the need for a quantum repeater to realize long-distance quantum communications\cite{pirandola2017fundamental}.

As of the time of writing, there have been more than a dozen related review papers summarized by a recent review of quantum repeaters~\cite{azuma2023}.   

The first quantum repeater concepts are based on entanglement distribution, purification and quantum teleportation~\cite{Briegel98,Dur99}. Here quantum communications can be achieved by distributing entanglement and performing quantum teleportation~\cite{bennett1993teleporting} with the distributed shared entanglement.  To deal with the nondeterministic throughput due to transmission loss, quantum memory is used to hold successfully distributed entangled resources until they are needed for quantum teleportation. To reduce the impact of exponentially increasing fiber loss, the channel is broken into smaller, lower-loss links. 

To mitigate the decoherence of quantum storage, entanglement purification protocols can transform a number of imperfect input entangled resources into a fewer number of higher quality entangled resources for quantum teleportation.   

\begin{figure}[t]
\includegraphics[ angle=-90,origin=c,width=0.95\textwidth]{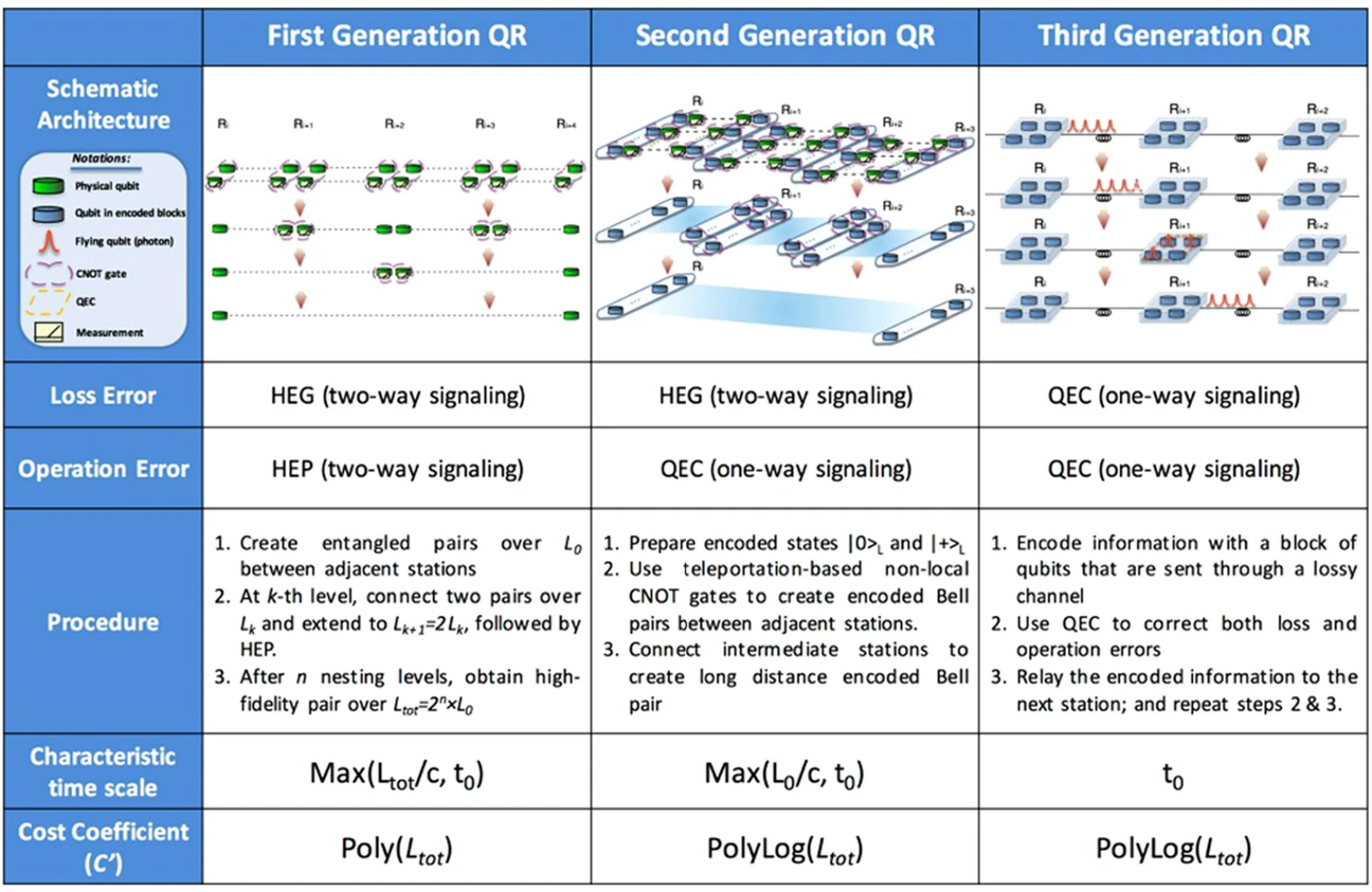}
\vspace{-2.5cm}
    \centering
    \caption{Quantum repeater protocols by generation. 
 Image reproduced from~\cite{Muralidharan16} under a Creative Commons Attribution 4.0 International License (\url{http://creativecommons.org/licenses/by/4.0/}).  Key steps include heralded entanglement generation (HEG), heralded entanglement purification (HEP), and quantum error correction (QEC). } 
	\label{fig:srep}
\end{figure}

 To capture the major differences in the development of  quantum repeater (QR) protocols, three generations have been identified by Munro and colleagues~\cite{Munro15} as well as by Muralidharan and colleagues~\cite{Muralidharan16}. As characterized in Fig.~\ref{fig:srep}, the different generations of QRs may be thought of by how they handle loss and operational errors, as well as the required classical communications (i.e., one-way or two-way signaling).    For first and second generation QRs, loss errors are corrected via heralded entanglement generation (HEG), while third generation QRs utilize quantum error correction (QEC).  For operational errors, heralded entanglement purification (HEP) is the exclusive domain of first generation QRs, while second and third generations use QEC.

 Experimental implementations of quantum repeaters still lag behind theoretical advances. To date, most demonstrations have been limited to entangling a small number of distant quantum memories---a key milestone toward realizing first-generation quantum repeaters~\cite{Hensen2015,lago2021telecom,vanLeent2022,Knaut2024}. While scalable, practical quantum repeaters remain out of reach, the technologies developed so far have already shown promise for near-term applications, such as device-independent quantum key distribution (DI-QKD)~\cite{Zhang2022}, which can establish security without requiring trust in most components of the QKD system. Furthermore, it has been shown that even a single repeater node can enhance communication rates: for instance, the secret key rate in QKD can surpass the fundamental limit of repeaterless protocols, improving from a linear $\eta$ scaling to a more favorable $\sqrt{\eta}$ scaling, where $\eta$ denotes the channel transmittance~\cite{bhaskar2020experimental,pu2021experimental}. 

While the development of QRs is a very active area of research, their integration into classical networks is likely to be a strong function of the underlying approach.  For example, some challenges first identified in the context of ``coexistence'' of classical and quantum channels for QKD, such as Raman scattering, may drive some QR concepts towards using dedicated dark fiber for the quantum transmission, while others may coexist. Either way, it seems likely that some classical-level signals for quantum channel control and stabilization will be needed even if it is advantageous to refrain from using remaining bandwidth to send additional classical data.

\subsubsection{Near-term application---quantum key distribution}

As highlighted in earlier sections, quantum repeaters are essential for enabling the reliable transmission of quantum information across lossy communication channels. However, since practical quantum repeaters remain far from realization, near-term quantum communication applications are likely to rely on protocols that do not require deterministic state transmission. A prominent example is QKD~\cite{bennett2014quantum, ekert1991quantum}, which allows two remote parties (commonly referred to as Alice and Bob) to establish random keys using an insecure quantum channel and a classical authenticated channel. The generated keys can then be applied to various (classical) cryptographic protocols where symmetric keys are required. For a recent review on QKD, see ~\cite{xu2020secure}.

The first QKD protocol, the celebrated BB84~\cite{bennett2014quantum}, was proposed by Bennett and Brassard at a time when it was assumed that the key distribution problem in cryptography was already solved by using public-key cryptosystems, such as the RSA algorithm ~\cite{rivest1978method}. Potential real-life applications of QKD were not recognized until quantum researchers realized that quantum computing might pose a formidable challenge to contemporary public cryptographic protocols, including RSA,  as evidenced by Shor’s algorithm for factoring ~\cite{shor1994algorithms}. This spurred the rapid growth of QKD and related quantum cryptographic protocols, such as quantum secret sharing (QSS) ~\cite{hillery1999quantum,cleve1999share,karlsson1999quantum}, quantum digital signatures (QDS) ~\cite{gottesman2001quantum,dunjko2014quantum}, etc. Today, commercial QKD systems are readily available, and several large-scale QKD networks have been deployed worldwide ~\cite{elliott2002building,Peev2009,Chen2009,Sasaki2011,Lopez2020,Evans2021}.

In a typical prepare-and-measure QKD protocol, Alice sends Bob a single copy of a photonic quantum state encoded with truly random numbers in each round of quantum transmission. These random numbers can be generated using a quantum random number generator (QRNG)~\cite{ma2016quantum,herrero2017quantum}. According to the quantum no-cloning theorem~\cite{Wootters1982,dieks1982communication}, any attempt by an adversary (Eve) to gain information about the encoded random numbers will inevitably disturb the quantum states sent by Alice. By communicating over an authenticated classical channel, the two legitimate users can estimate the noise introduced in the quantum channel and use it to place an upper bound on Eve's information. If this information remains below a certain threshold, Alice and Bob can proceed to generate a final secret key by performing \textit{classical} error correction and privacy amplification ~\cite{bennett1988privacy,bennett1992experimental}.

In the QKD process described above, random numbers---rather than meaningful messages---are transmitted during the quantum communication stage. A shared random key can be generated as long as some of the transmitted random numbers are successfully detected by Bob. This implies that QKD does not require deterministic quantum communication and can be implemented without the need for quantum repeaters. Moreover, error correction and privacy amplification are carried out during the classical communication stage, and no quantum error correction is required. Thanks to the above features, the implementation of QKD is much less demanding than most other quantum communication protocols.

In classical optical communication, the two most commonly used detection schemes are direct detection and coherent detection (cf. Fig.~\ref{fig:3}). Their counterparts in QKD are single-photon detection and optical coherent detection, respectively. QKD protocols are generally categorized into two families: DV-QKD and CV-QKD~\cite{xu2020secure}, based on the dimensionality of the encoding space. However, the defining characteristics of different QKD protocols are determined primarily by the detection scheme used. Table 2 summarizes the main features of several practical QKD protocols. For more details on practical implementations of QKD and QKD networks, readers are referred to recent review papers~\cite{xu2020secure,cao2022evolution}.

%Table two--Practical QKD protocols
\begin{table}[ht]
\centering
\footnotesize
\caption{Some Practical QKD Protocols}
\label{tab:QKDprotocol}
\begin{tabular}{@{}|p{3cm}|p{11.5cm}|@{}}
\hline
Protocols & Main Features \\
\hline
Decoy-State BB84 QKD~\cite{hwang2003quantum,lo2005decoy, wang2005beating} & 

(a) The decoy-state scheme enables the implementation of single-photon-based protocols using conventional laser sources without significantly compromising performance. Decoy-state BB84 QKD is one of the most extensively studied and widely deployed DV-QKD protocols.

(b) A secret key rate of 13.72 Mb/s has been demonstrated over a 2~dB channel using InGaAs avalanche photodiode detectors~\cite{yuan201810}, and a key rate of 115.8~Mb/s over 10~km of standard optical fiber has been achieved using multipixel superconducting nanowire single-photon detectors~\cite{li2023high}.

(c) Using a three-state time-bin protocol combined with a one-decoy approach, secure QKD over a distance of 421~km has been demonstrated using ultralow-loss fiber~\cite{boaron2018secure}. Remarkably, satellite-to-ground QKD has also been demonstrated over a 1200km free-space channel~\cite{liao2017satellite}.
 \\
\hline
Gaussian-Modulated Coherent States (GMCS) QKD \cite{grosshans2003quantum} &

(a) The Gaussian-modulated coherent-state (GMCS) QKD protocol, based on conventional lasers, modulators, and optical homodyne detectors, is one of the most extensively studied CV-QKD schemes and is considered a cost-effective solution for practical deployment.

(b) The mode-selection function of the LO in coherent detection makes GMCS QKD particularly attractive for classical and quantum co-existence networks~\cite{qi2010feasibility}.

(c) The room-temperature operation of optical homodyne detectors paves the way for on-chip QKD implementations~\cite{zhang2019integrated}, especially when the LO is generated at the receiver’s end~\cite{qi2015generating,soh2015self}.

(d) Without detector deadtime, GMCS QKD can achieve very high secret key rates over low-loss channels. Recently, using a passive-state-preparation scheme~\cite{qi2018passive}, the authors in~\cite{ji2024gbps} estimated a gigabit-per-second secret key rate could be achieved over short distances.

(e) Due to the unavoidable presence of vacuum noise, GMCS QKD typically operates in a very low SNR regime, making it less suitable for channels with high loss. The current maximum distance achieved over ultralow-loss fiber is 202~km~\cite{zhang2020long}.
  \\
\hline
Measurement-device-independent (MDI) QKD~\cite{lo2012measurement} & 

(a) Measurement-device-independent (MDI) QKD enables secret key distribution without requiring trust in the measurement device, rendering it inherently immune to all detector side-channel attacks.

(b) MDI-QKD networks also allow multiple users to share a centralized, high-cost detector, and have been successfully demonstrated over deployed optical fiber networks~\cite{tang2016measurement}. 

(c) Several variants of MDI-QKD, including twin-field QKD~\cite{lucamarini2018overcoming} and mode-pairing QKD~\cite{zeng2022mode,xie2022breaking}, improve the key rate scaling from linear with channel transmittance $\eta$ to a square-root dependence $\sqrt{\eta}$, leading to a recent demonstration of QKD over 1000~km of ultralow-loss fiber~\cite{liu2023experimental}.
 \\
\hline
\end{tabular}
\end{table}

Single-photon detectors are commonly employed to implement DV-QKD, where the dimension of the encoding space (such as the polarization of a single photon) is finite, although they have also been applied in certain CV-QKD protocols, such as time-frequency-coded QKD~\cite{qi2006single,qi2011quantum,nunn2013large,mower2013high}, where the dimension of the encoding space is infinite. In contrast, optical coherent detectors are almost exclusively used in CV-QKD protocols based on quadrature encoding using coherent states or squeezed states~\cite{ralph1999continuous,hillery2000quantum,grosshans2003quantum}. An interesting variation is operating a coherent receiver in phase-insensitive photon detection mode~\cite{qi2020characterizing} to implement DV-QKD protocols~\cite{qi2021bennett}. Such an approach may inherit certain advantages of both detection schemes.

To make QKD practical, considerable efforts have been devoted to enhancing the security of real-world systems and reducing implementation costs. While the security of an idealized QKD protocol can be rigorously analyzed based on the fundamental laws of quantum physics, real-world implementations are inevitably imperfect. An adversary may exploit overlooked weaknesses in the physical implementation through side-channel attacks~\cite{zhao2008quantum,lydersen2010hacking,xu2010experimental}. Although completely eliminating all potential side channels may not be feasible, practical security can be strengthened by minimizing the attack surface. In this context, various device-independent (DI) and semi-DI QKD protocols have been proposed, where security can be guaranteed even when some QKD devices are untrusted or uncharacterized~\cite{mayers1998quantum,acin2007device,zapatero2023advances}. Among them, the measurement-device-independent (MDI) QKD protocol~\cite{lo2012measurement} stands out for its great balance between security and feasibility.

QKD is fundamentally a point-to-point protocol. To fully realize its potential, large-scale QKD networks with broad geographical coverage, accessible to a large number of users, are required. However, in the absence of QRs, there are inherent limitations on the achievable quantum communication rates over a lossy bosonic channel. This understanding emerged from the historical effort to determine the secure key rate achievable using existing QKD protocols. The fundamental rate-loss tradeoff was first established by Takeoka, Guha, and Wilde in the context of secret key agreement over a noisy lossy channel, leading to the ``TGW'' bound~\cite{takeoka2014fundamental}. Subsequently, Pirandola, Laurenza, Ottaviani, and Banchi expanded the analysis to cover a broader range of quantum channels, culminating in the ``PLOB'' bound for quantum repeaterless channels~\cite{pirandola2017fundamental}.

One solution to surpass the PLOB bound, which dictates that the secret key rate scales linearly with the channel transmittance $\eta$, is to divide the quantum channel into smaller segments by introducing intermediate ``trusted" nodes along the path, allowing for QKD between adjacent nodes in parallel. Once a random key is established between any pair of adjacent nodes, Alice’s key can be transmitted to Bob hop-by-hop using a classical one-time-pad encryption scheme~\cite{elliott2002building}. The secret key rate in this scheme is limited only by the loss of the longest segment, rather than the loss across the entire link. However, since all intermediate nodes will possess a copy of the secret key, the security of this approach depends on the trustworthiness of every intermediate node. Another method to surpass the PLOB bound is to utilize specialized variations of MDI-QKD, such as twin-field QKD~\cite{lucamarini2018overcoming} or mode-pairing QKD protocols~\cite{zeng2022mode,xie2022breaking}, which improve secret key rate scaling from $\eta$ to $\sqrt{\eta}$. However, it remains unclear how to extend this approach to involve more than one intermediate node. Recently, QKD over 1000 km of low-loss fiber has been demonstrated using the twin-field QKD protocol~\cite{liu2023experimental}.

Similar to classical networking, an efficient many-user network in quantum communication depends on a well-designed switching (or routing) scheme. Most existing QKD networks are structured like ``circuit-switched'' networks in classical communication: before two remote parties can begin QKD, a direct optical link must be established and reserved for the duration of the quantum communication phase. More recently, packet switching, which has been shown to offer better efficiency in classical networking, has been introduced into quantum communication and QKD~\cite{diadamo2022packet,mandil2023quantum}.

To enable packet switching of quantum signals, hybrid frame structures are likely required. In these structures, each quantum payload is encapsulated by a classical header, which provides essential information for routing, error mitigation and correction. Unlike the classical header, which can be measured and regenerated at each network switch, the quantum payload must remain in the optical domain. This approach closely parallels the concept of optical packet switching in classical networking and consequently faces similar challenges.

\begin{figure}[t]
	\includegraphics[width=.8\textwidth]{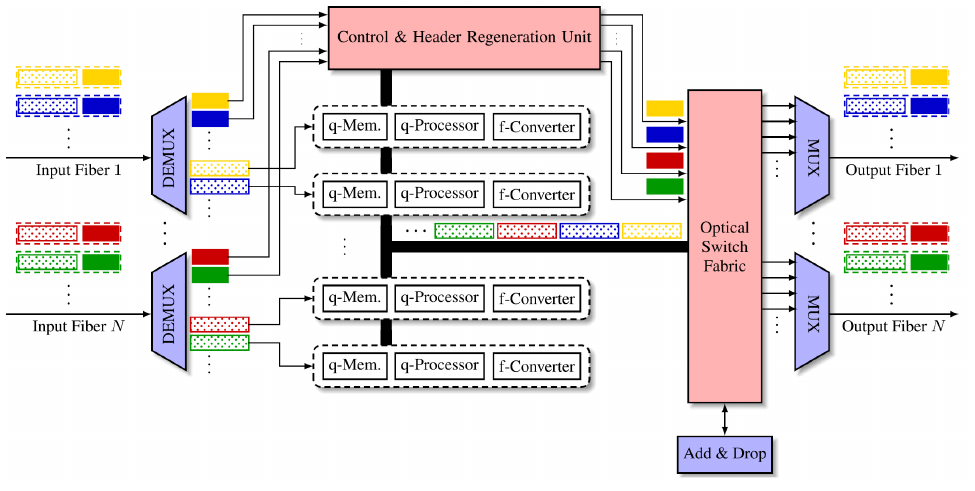}
    \centering
    %%	\captionsetup{justification=raggedright, singlelinecheck=false }
	\caption{A conceptual design of a quantum ROADM. The classical header can be measured and regenerated in the classical processing unit, while the quantum payload is stored, corrected for errors, and, if needed, undergoes frequency conversion in the quantum processing unit.  Image reproduced with permission from a Creative Commons Attribution 4.0 International License (\url{http://creativecommons.org/licenses/by/4.0/})~\cite{diadamo2022packet}.} 
	\label{fig:PS}
\end{figure}

A critical component of packet-switched quantum networks is the quantum counterpart of the reconfigurable optical add-drop multiplexer (ROADM), known as the q-ROADM, illustrated in Fig.~\ref{fig:PS}. Although it is too early to present a detailed design, developing a general-purpose q-ROADM will demand significant advancements in fundamental quantum technologies, including long-lived quantum memories and QRs.

\subsubsection{Quantum communication networks}
\label{sec:quantumMultiuser}
As in classical networking, the basic qualification of any quantum \emph{network}---as opposed to a quantum \emph{link}---is the support of $N>2$ users. The implementation of such multiuser support can prove as simple as connecting a central source to $N$ nodes through a passive $1\times N$ coupler. Indeed, in the first multiuser quantum communications experiment, a single source of weak coherent light was probabilistically split among three receivers, demonstrating the ability to perform QKD independently between one Alice and multiple Bobs~\cite{Townsend1997a}. Broadcasting through passive splitters can be extended to entangled light sources and represents an effective solution for fixed QKD networks, yet it suffers from significant limitations for quantum networking more generally in that it provides no bandwidth management or reconfigurability and is lossy intrinsically (rather than just technically): i.e., no two users can be connected through a direct lightpath, even temporarily, without additional scattering to other fibers in intervening couplers.

In view of such limitations, multiplexing strategies based on classical lightwave communications (cf. Sec.~\ref{sec:multiuser}) have been increasingly adopted in the quantum domain. As general guidelines, any prospective ``quantum-capable'' multiplexing approach should be (i)~all-optical in the broad sense (i.e., implementable without OE conversion) and (ii)~avoid amplification (i.e., amplification must not be intrinsic to its operation). Standard OE conversion represents a measurement that collapses the quantum state, and qubits cannot be amplified without introducing noise~\cite{Wootters1982,dieks1982communication}. Thankfully, WDM does meet these requirements and %offers an expansive infrastructure for quantum multiplexing
provides a powerful toolkit for allocating quantum resources.

In particular, the frequency entanglement generated through narrowband-pumped spontaneous parametric downconversion~\cite{Shih2003} offers broadband spectral correlations that serve as natural resources for entangling many users; for example, when combined with entanglement in another DoF such as polarization or time bins, pairs of frequency-entangled bands can be allocated to any two parties wishing to share entanglement. This approach was demonstrated 
in a fully connected four-user QKD experiment facilitated by a shared polarization---frequency hyperentangled photon source and a network of interconnected DWDM filters: first to demultiplex twelve 100~GHz-wide bands, and then to multiplex channel triplets destined for each receiver~\cite{Wengerowsky2018}. In analogy with classical WDM, this experiment can be viewed as a realization of ``fixed DWDM,'' where the allocation is static and each channel occupies a uniform spectral slot. 

More recent experiments have expanded on this concept by bringing elastic optical networking concepts squarely into the quantum domain; replacing fixed DWDMs with a single WSS was shown to enable multiuser entanglement distribution in which each pair of users is assigned bandwidth dynamically and in proportion to demand~\cite{Lingaraju2021, Appas2021}. Although WSSs impart higher losses ($\sim$5~dB) compared to a single DWDM ($\sim$0.5~dB), they can support nodes without additional components, thus offering better scaling in the high-user regime~\cite{Lingaraju2021}. 
For example, a recent experiment merging a broadband Sagnac polarization-entangled source~\cite{Alshowkan2022c} with C- and L-band WSSs demonstrated high-fidelity entanglement in 150 pairs of 25~GHz-wide channels, which when combined with the high port counts (e.g., $\geq$20) and fine spectral granularity (e.g., 3.125~GHz) available from modern WSSs~\cite{Ma2021}  point to a promising future for flex-grid quantum networking.

\begin{figure}[t]
	\includegraphics[width=.8\textwidth]{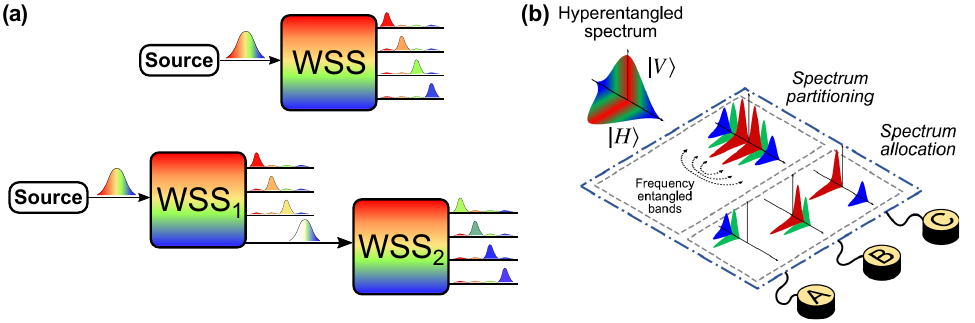}
    \centering
    %%	\captionsetup{justification=raggedright, singlelinecheck=false }
	\caption{Flex-grid entanglement distribution with wavelength-selective switches (WSSs). (a)~Network  expansion by nesting multiple WSSs~\cite{Alshowkan2021}. (b)~Distribution of polarization-frequency hyperentangled bands to realize a fully connected entanglement network~\cite{Alnas2022}. Images reproduced with permission from a Creative Commons Attribution 4.0 International License (\url{http://creativecommons.org/licenses/by/4.0/})~\cite{Alshowkan2021} and an Optica Publishing Group Open Access License~\cite{Alnas2022}.} 
	\label{fig:J1}
\end{figure}

In such a future, one can envision leveraging multiple WSSs connected together to support complex mesh or multihop network topologies. Figure~\ref{fig:J1}(a) highlights one simple example, whereby a source of frequency-entangled photons distributed by a single WSS in a star configuration (top) is expanded to service a second set of nodes of with a second WSS (bottom). Assuming these WSSs are associated with distinct quantum local area networks (QLANs), for example, this construction represents perhaps the simplest quantum internetwork---i.e., a true ``quantum Internet''~\cite{Alshowkan2021}. Depending on the design of the optical source, quantum information can be carried in any DoF compatible with the fiber transmission medium; Fig.~\ref{fig:J1}(b) depicts a version in which three separate pairs of polarization-frequency hyperentangled bands (red, green, and blue) are allocated to supply polarization entanglement to all pairs of users in a three-node network~\cite{Alnas2022}.

More broadly, WSSs form just one example---albeit a particularly sophisticated one---of a variety of spatial switching technologies available for quantum networking, including microelectromechanical and optomechanical versions~\cite{Toliver2003, Peters2009, Chapuran2009, Herbauts2013, Laudenbach2020}. In all cases, the reconfigurability available permits dynamic allocation of bandwidth to pairs of users on demand, facilitating any-to-any connectivity without the dedicated wavelengths and extra background light  inherent to a fully connected quantum network in the fixed DWDM paradigm~\cite{Alshowkan2021, Alnas2022}.

Complementary to the variety of WDM approaches available and already demonstrated in quantum networking, SDM is beginning to emerge in tabletop demonstrations with multicore fiber, mirroring this evolving field in classical fiber optics~\cite{Dynes2016, Ding2017, Lee2017, Canas2017, Xavier2020, Gomez2021, Ortega2021, Achatz2023}. Extending even further back in the history of quantum communications, TDM techniques have likewise made their impact felt. In experiments so far, it is useful to classify TDM as either ``slow'' or ``fast,'' where the former refers to temporal segmentation of network resources into stretches of time much longer than the symbol period, and the latter to time divisions at the clock rate for multiplexing users at the symbol-by-symbol level. Slow TDM therefore encompasses arrangements in which quantum communication sessions alternate with classical ones---valuable for tracking drifts in polarization~\cite{Treiber2009} or interferometric phase~\cite{Humphreys2018} or for reducing Raman noise during quantum transmissions~\cite{Choi2011}. Although less common, fast TDM has proven extremely effective in CV-QKD, where the interleaving of bright pilot pulses with few-photon quantum signals enables phase recovery with a truly ``local''---and hence secure---LO~\cite{qi2015generating, soh2015self}.

With the multiplexing (demultiplexing) occurring prior to modulation (after detection), this CV-QKD example represents \emph{electrical} TDM in the definition of Sec.~\ref{sec:multiuser}; as in the classical domain, the switching requirements for fast \emph{optical} TDM have prevented significant application in practical links. But whereas packet switching---the next level of temporal resource management---can be realized classically in the electronic domain, only optical packet switching is available for quantum signals, since any packet must be switched without destroying the quantum information carried. Quantum packing switching architectures have been proposed and show that, in principle, more efficient resource utilization is possible~\cite{diadamo2022packet, mandil2023quantum, Yoo2024}, with a recent experiment demonstrating $\upmu$s-scale switching of single-photon signals based on classical headers~\cite{On2024}. Indeed, it is difficult to imagine a quantum Internet of the same scale as the classical Internet \emph{without} packet switching, yet the technological challenges make it unclear where the cost-benefit tradeoff will land. It is certainly possible that the circuit switching which has come to dominate the optical physical layer in conventional networks will persist in quantum networks but at higher networking layers. The applications supported by a quantum wide-area network of scale comparable to traditional telephony, the archetypal circuit-switched network, would be inconceivably more powerful than the quantum information processing available today---whether or not it scales like the Internet.

Nonetheless, it appears to us that the fundamental requirements of quantum networking furnish the strongest motivation to date for developing all-optical signal processing approaches. Indeed, both quantum frequency conversion and quantum transduction~\cite{Kumar1990, Lauk2020, Awschalom2021, Han2021} will be critical for interfacing stationary quantum systems with telecom-band photons, required irrespective of whether circuit or packet switching is employed. And an extensive toolkit for quantum gates in SMF-compatible DoFs like time bins~\cite{Humphreys2013, Takeda2019}, frequency bins~\cite{Kues2019, Lu2019c, Pfister2020, Lu2023a, Lu2023c}, and orthogonal modes~\cite{Brecht2015, Ansari2018, Raymer2020, Karpinski2021}---as well as MMF-compatible DoFs like transverse spatial modes~\cite{Defienne2016, Valencia2020}---has emerged in recent years.
Although experiments so far have demonstrated local quantum information processing functionalities rather than distributed networking per se, the nonlinear, electro-optic, and pulse shaping techniques employed have significant potential in this space; for example, the frequency-bin discrete Fourier transform and interleaved beamsplitter---both recently experimentally realized via the electro-optic-based quantum frequency processor (QFP)~\cite{Lu2022a, Lingaraju2022}---form potential building blocks for spectrally multiplexed quantum interconnects~\cite{Lu2023c}. The degree to which these spectro-temporal and spatial DoFs will  assist not only in multiplexing quantum information but also in encoding the information itself remains an open question and fertile ground for continued research.

\section{Integration of quantum communications in conventional fiber-optic networks}
\label{sec:integration}
Early quantum communication experiments were largely conducted over ``dark fiber'', where no classical communication signals were present alongside quantum signals. The rationale is clear: classical signals are many orders of magnitude more intense than quantum signals, and even minor leakage from classical signals can severely degrade quantum signals. However, dark-fiber links are scarce and expensive. Given the limited near-term applications of quantum networking, enabling the coexistence of quantum and classical traffic within standard WDM fiber networks offers a promising route to commercialization. To this end, significant research has focused on enhancing the compatibility of quantum communication protocols with existing fiber-optic infrastructures. Table \ref{tab:Coexistence} summarizes the major challenges associated with transmitting quantum signals over classical fiber-optic networks, along with potential mitigation strategies. A detailed review of the coexistence of QKD and classical traffic in the same fiber can be found in~\cite{cao2022evolution}.

%Table three--Key challenges in transmitting quantum signals through classical fiber-optic networks and possible mitigation strategies

\begin{table}[ht]
\centering
\footnotesize
\caption{Transmitting quantum signals through classical fiber-optic networks.}
\label{tab:Coexistence}
\begin{tabular}{@{}|p{5cm}|p{9cm}|@{}}
\hline
Challenges & Possible mitigation strategies \\
\hline
Noise photons generated by intense classical traffic can significantly degrade the SNR of quantum communications (including crosstalk and scattered conventional signal noise). & 

(a)	Optimize wavelength assignment in WDM networks to minimize crosstalk~\cite{Townsend1997b}.

(b)	Apply additional spectral and temporal filtering---such as narrowband optical filters and detector gating---to suppress noise photons in quantum channel~\cite{Peters2009}.

(c)	Leverage the intrinsic mode-selectivity of optical homodyne detectors in coherent quantum communication to reject out-of-mode noise~\cite{qi2010feasibility}.

(d)	Use novel fiber types, such as hollow-core fibers, which exhibit reduced nonlinearity and scattering effects, thereby minimizing classical-to-quantum channel interference~\cite{alia2022dv}.

(e) Use Procrustean local filtering to remove conventional signal crosstalk from distributed entanglement~\cite{Lu2024}.
 \\
\hline
Certain devices commonly used in classical optical networks---such as optical amplifiers, classical repeaters, and routers---are incompatible with quantum communications. &

(a)	Bypass devices that are incompatible with quantum signals ~\cite{Nweke06}. Note that such bypasses must block amplified spontaneous emission or other noise emitted by the bypassed device, including amplifiers and lasers~\cite{runser07},  in the quantum transmission band.

(b)	Develop and deploy hybrid network components, such as quantum-compatible routers, that can support both classical and quantum communications~\cite{diadamo2022packet}.
\\
\hline
Quantum communications are generally more sensitive to channel impairments and noise than classical systems. & Advance noise-reduction technologies—such as precise polarization control, dispersion management, synchronization, and optical phase recovery—which enhance the performance of quantum communication and can also benefit classical channels.
 \\
\hline
\end{tabular}
\end{table}

\subsection{Coexistence of quantum and classical signals in telecom fiber}
\label{sec:coex}
As discussed in Sec.~\ref{sec:quantumMultiuser}, classical WDM technology has proven extremely beneficial for resource allocation in multiuser quantum communications. %sharing of limited optical resources resources by many quantum users in the same fiber infrastructure. 
Yet the value of WDM for quantum networking does not stop at quantum signals alone; it also offers a means for sharing fiber resources with typically much brighter classical signals as well---so-called quantum-classical coexistence.

In a typical coexistence architecture based on WDM, noise photons in the quantum channel can originate from various sources. These include leakage photons from classical channels due to the finite isolation of multiplexing and demultiplexing components, in-band noise photons generated in optical fibers through nonlinear processes such as four-wave mixing (FWM) and spontaneous Raman scattering, and broadband ASE photons produced by optical amplifiers. While out-of-band noise photons can be mitigated by using high-quality components with better isolation, suppressing in-band noise requires additional measures, such as extra filtering in the quantum channel, careful wavelength allocation, and power control in classical channels. In this context, CV quantum communication protocols based on optical coherent detection are particularly promising due to the ``mode-selection'' capability of the LO. This mechanism allows only photons in the same spatiotemporal and polarization mode as the LO to be effectively detected, while noise photons in other modes are naturally filtered out.

Historically, the initial drivers for WDM in quantum communications centered on the coexistence of DV-QKD signals with classical traffic, with the first demonstration placing the former in the O-band (1260--1360~nm) and the latter in the C-band (1530--1565~nm)~\cite{Townsend1997b}. A quantum/classical O/C-band split has been the preferred allocation in most subsequent QKD coexistence experiments~\cite{Chapuran2009, Choi2011, Wang2017, Mao2018, Valivarthi2019, Gruenenfelder2021, Berrevoets2022}, although other CWDM channel combinations---such as quantum in the C-band and classical in L-band (1565--1625~nm)~\cite{Patel2012}, or quantum in the C-band and classical in the O-band, L-band, and S-band (1460--1530~nm)~\cite{Froehlich2015}---have been realized as well. Pushing even tighter to DWDM spacings has enabled positive QKD rates over tens of kilometers with copropagating classical channels as close as 200~GHz~\cite{Peters2009, Eraerds2010}, highlighting the feasibility of ultradense bandwidth allocations.

More recently, quantum-classical coexistence experiments have branched out beyond QKD to more general scenarios such as quantum process tomography~\cite{Chapman2023}, quantum state tomography~\cite{Thomas2023}, and quantum teleportation~\cite{Thomas2024}. Arguably \emph{the} defining impairment in all of these experiments is Raman scattering---noise photons generated through light interactions with vibrational modes in silica fiber~\cite{Stolen1984}. Related to both the classical launch power and fiber length, Raman is challenging to handle, since its spectral content can overlap with the quantum band of interest; the fact its intensity is slightly higher on the redshifted side of the signal and drops for large spectral separations explains the general experimental preference for O-band quantum signals paired with C-band classical ones, yet the DWDM experiments above~\cite{Peters2009, Eraerds2010, Chapman2023}---as well as nonlocal biphoton modulation with coexisting radio-over-fiber timing synchronization signals~\cite{Chapman2024opticaQ}---highlight the feasibility for C-band coexistence as well, where both signals can enjoy the lowest-loss region of optical fiber.

\subsection{Quantum incompatible devices in classical network and solutions}
\label{sec:quantumIncompatible}
Classical networks are carefully engineered such that the transmitted data signals are reliably received by the paired receivers.  This requires that there be sufficient SNR at the receiver to faithfully decode the transmitted classical information.  In addition, with current technology, it is common for classical receivers to be completely insensitive to quantum signals. Consider a quantum transmitter which emits a stream of perfect 1550~nm single photons (each about 0.8 eV) at 100 billion times a second.  Such a quantum transmitter, while beyond the realm of current experimental feasibility, would have an average launch power of roughly five orders of magnitude below that of a typically launched conventional signal (roughly 1 mW or 0 dBm).  In addition, classical network devices can emit optical noise that is below the sensitivity threshold of classical receivers. As a result, conventional network devices may not emit noise detectable by a classical receiver, yet they may emit significant noise for quantum networks, potentially impairing the function of a quantum receiver.     

One such device is an optical amplifier.  While in the context of no-cloning, an amplifier adds noise to amplified photons, it can also add noise to a quantum network without an input signal via ASE. ASE can have a broad bandwidth, and care must be take to prevent that noise from being introduced directly into a quantum network.   For example, the most common types of networked amplifiers, EDFA, have ASE bandwidths from 1530--1565 nm. Other similar amplifiers include semiconductor optical amplifiers (SOAs), which also emit ASE noise.  Semiconductor lasers can emit noise at much higher bandwidths than the emission wavelength albeit at a much lower intensity than the fundamental emission wavelength.  If the spontaneous emission overlaps with the quantum signal wavelength(s), that part of the spontaneous nose spectrum may need to be filtered out before the spectrum is combined onto a fiber with a quantum signal.  In recognition of these noise sources, an early experiment used an amplifier for mid-span amplification of conventional signals in the C-band, while bypassing an O-band quantum signal~\cite{Nweke06}.  In addition, both the quantum signal and conventional laser light were filtered to reduce noise in early QKD coexistence experiments~\cite{Peters2009}.

\subsection{Simultaneous transmission of classical and quantum information}

Replacing the existing optical fiber infrastructure is prohibitively expensive. Therefore, in the foreseeable future, the most promising quantum communication applications will likely be those that are compatible with current fiber networks. As noted in earlier sections, significant progress has been made in developing loss-tolerant protocols, such as QKD, that can operate over conventional optical fibers alongside classical traffic~\cite{Peters2009,qi2010feasibility,kumar2015coexistence,eriksson2019wavelength,Mao2018}. By allowing multiple applications to share the same optical fiber, the implementation cost of quantum communication protocols can be substantially reduced.

A significant next step in this direction is the shared use of network resources, including both hardware and software, for classical and quantum communications. In particular, CV-QKD is especially attractive because its information encoding and decoding methods, as well as the required hardware, are remarkably similar to those of classical coherent optical communication. Recent advancements in both fields have further reduced the gap between them. On the classical side, continuous improvements in detector performance and the development of forward error correction (FEC) have enabled bit error rates (BER) as low as $10^{-9}$ with just a few photons per bit~\cite{stevens2008optical}. This implies that state-of-the-art coherent communication systems can already operate in the quantum noise-dominant regime. On the quantum side, new security proof technologies in CV-QKD~\cite{ghorai2019asymptotic,lin2019asymptotic} allow much-simplified encoding schemes, such as binary phase-shift keying (BPSK) and quadrature phase-shift keying (QPSK), which are widely used in classical communication. Furthermore, novel phase recovery methods~\cite{qi2015generating,soh2015self}, now enable CV-QKD to use a true LO generated at the receiver, eliminating the need for a transmitted LO from the sender. This local LO scheme has been experimentally demonstrated over 100~km of optical fiber~\cite{hajomer2024long}.

An active area of research involves leveraging advanced classical coherent communication systems, along with their corresponding DSP technologies, to implement quantum communication protocols. As the hardware requirements for classical and quantum communication converge, it becomes increasingly feasible to use the same infrastructure for both, operating in a time-sharing manner. Such a universal coherent communication system could play a pivotal role in the next-generation fiber-optic communication networks~\cite{qi2024hybrid}.

Classical information can generally be considered as a special case of quantum information. Therefore, a quantum communication system should, in principle, be capable of transmitting both quantum and classical information. The capacity of a quantum channel for simultaneously transmitting these two types of information was explored in~\cite{devetak2005capacity}, where the authors established a general trade-off relationship between quantum and classical capacities. Further trade-off relations, specifically for public communication, private communication, and secure key generation, were derived in~\cite{wilde2012information}, assuming no feedback from the receiver to the sender.

Recently, a simultaneous classical communication and QKD protocol was proposed in~\cite{qi2016simultaneous}. The key idea is to superimpose appropriately scaled random QKD signals onto the strong classical information signals and resolve both at the receiver through optical coherent detection. For classical communication, the QKD signals manifest as noise, reducing the SNR. Nevertheless, as long as the modulation variance of the QKD signals remains sufficiently small, the bit error rate for classical communication can be effectively mitigated using FEC~\cite{chang2010}, which is widely employed in modern optical systems to ensure high data transmission accuracy with minimal optical power. After reliably recovering the classical information, the receiver can digitally subtract it from the measurement results to extract the encoded QKD signals. Notably, this protocol has a stealth feature akin to covert communication~\cite{bash2015quantum}: legitimate users can introduce additional noise even when not conducting QKD, making it difficult for an adversary to detect whether QKD is in progress.

\subsection{Beyond the physical layer}
In the existing Internet, the Transmission Control Protocol/Internet Protocol (TCP /IP) stack~\cite{Cerf1974, Braden1989} has provided an enduring architecture for facilitating communications on a global scale. Classifying protocols and hardware into five distinct layers---physical, link, network, transport, and application---the TCP/IP stack facilitates both abstraction and modularity: each layer needs only to concern itself with the services for which it is responsible, with the freedom to implement intralayer hardware and software modifications without impacting those below or above. The success of the TCP/IP stack suggests the long-term importance of a \emph{quantum} Internet stack as well, by whose abstraction users would be able to employ quantum networks for sophisticated purposes in sensing, computing, and communication without the need to understand their intricacies of operation.

Yet despite the potential value of a quantum Internet stack, nearly all quantum communications research so far has been confined to what the TCP/IP stack would place in the physical layer, namely, protocols for and transmission of quantum information over point-to-point links. Of course, the unique features of quantum communications derive from the physical properties of quantum systems, so the historical emphasis on the physical layer has not only been reasonable, but required to push the field forward. However, with growing interest in more general multinode quantum networks, several quantum stacks analogous to TCP/IP have been proposed. The quantum recursive network architecture represents one of the earliest formulations with explicitly quantum layers~\cite{VanMeter2013}. Other proposed stacks focus on the establishment of multipartite graph states~\cite{Pirker2019}, distinguish layers by the increasing reliability and distance of entanglement generation~\cite{Kozlowski2019, Dahlberg2019, Kozlowski2020}, or identify layers with functionalities in wavelength-multiplexed entanglement distribution~\cite{Alshowkan2021, Alshowkan2022a}. The latter two stacks have even been implemented experimentally in deployed testbeds with nitrogen-vacancy (NV)~\cite{Pompili2022} or photonic polarization~\cite{Alshowkan2021} qubits, respectively. 

In light of the continued evolution of quantum networking technology, it remains unclear what quantum Internet stack---if any---will ultimately assume a place as ubiquitous as the TCP/IP stack in the classical Internet. Indeed, it is possible that attempting to align solutions with TCP/IP, even qualitatively, may artificially constrain and eliminate quantum network architectures from consideration that might prove better suited to the physical limitations of available hardware. Accordingly, at this stage it would appear prudent to develop quantum stacks in tune with \emph{both} TCP/IP-like abstraction models \emph{and} the performance of quantum devices like sources, memories, and repeaters; in this way, best practices from classical networks can be incorporated into quantum networks while avoiding physical-layer demands that might be difficult to satisfy with available quantum resources (e.g., indefinitely long storage times in quantum memories or unit-fidelity sources of entanglement).

One conventional network trend particularly apropos for quantum networks is software-defined networking (SDN), a reconfigurable architecture for network management in which control and data traffic are separated~\cite{Kim2013, Feamster2014, Kreutz2015}. Formally, SDN delineates three separate planes: (i) a data plane containing network equipment and access points, (ii) a control plane comprising SDN controllers that compute and communicate forwarding tables to data plane devices, and (iii) an application plane that communicates network requirements to the controllers for implementation. Whereas the separation of control and data planes in classical networking is motivated by pragmatism, such separation is arguably inherent in quantum networks, where the data of interest reside in quantum states that are fundamentally distinct from classical information in control and management signals. Consequently, SDN's explicit division of data and control planes provides a readily translatable framework for quantum networks, with several proposals in the literature~\cite{Aguado2017, Humble2018, Aguado2019, Alshowkan2022a, Chung2022}.

\begin{figure}[t]
	\includegraphics[width=.6\textwidth]{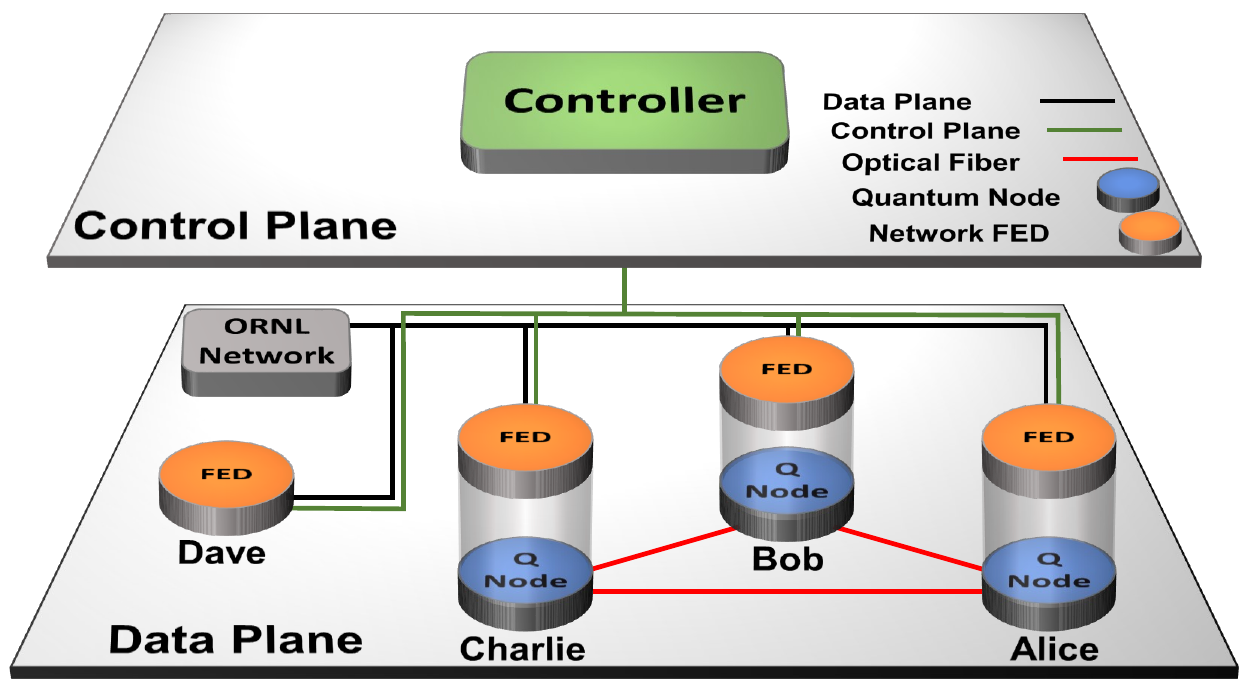}
    \centering
	\caption{Software-defined networking (SDN) architecture implemented on the Oak Ridge National Laboratory (ORNL) quantum network. The control plane manages firewall and encryption devices (FEDs) for classical communications, as well as quantum-optical hardware for entanglement distribution. Image reproduced with permission from Springer Nature~\cite{Alshowkan2022a}.} 
	\label{fig:J2}
\end{figure}

As a concrete example, Fig.~\ref{fig:J2} depicts control and data planes for a deployed QLAN leveraging a quantum-conventional network (QCN) harness architecture over three quantum-capable nodes (Alice, Bob, Charlie) and one classical node (Dave)~\cite{Alshowkan2022a}. The data plane consists of both classical and quantum traffic, the former comprising digital signals over ethernet and the latter photonic polarization qubits over direct optical fiber paths. The SDN controller in the control plane manages both (i)~firewall and encryption devices (FEDs) establishing secure tunnels for internode classical communications and (ii)~components in quantum nodes such as WSSs that define the spectral content of each lightpath. Compared to conventional SDN, the only significant addition is the quantum data plane, which can be managed directly by the control plane via TCP/IP communications. Moreover, although the current implementation relies on circuit switching in the quantum data plane, it would readily adapt to quantum packet switching as well; in that case the controller would transmit lookup tables to quantum switches and routers, rather than bandwidth and port assignments to WSSs. 

The ways in which packet and circuit switching impact SDN highlight the generally important relationship between SDN and the adopted quantum Internet stack. For just as classical SDN provides resources for managing a data plane in the TCP/IP stack, the types of routing decisions made by a quantum-conventional SDN controller will depend on the architecture of the quantum data plane; the stacks proposed above offer provisional examples for continued research~\cite{VanMeter2013, Pirker2019, Kozlowski2019, Dahlberg2019, Kozlowski2020, Alshowkan2021, Alshowkan2022a}. Therefore while the natural synergy of SDN with quantum networking sparks optimism in the path toward integrating quantum and classical networks on large scales, much work remains---both in fleshing out detailed protocols and in developing the technology capable of realizing them.

\subsection{Quantum network demonstrations}
\label{sec:qNetDemo}
The ultimate goal of any network---quantum or classical---is to enable communications between distributed users, and thus networks are often classified by the spatial scale of the nodes serviced: e.g., local, metro, and wide area networks (LANs, MANs, and WANs, respectively) and their quantum equivalents (QLANs, QMANs, and QWANs). Apart from the specific application of connecting modular quantum computers in a single laboratory, quantum networks by their very nature must move from the laboratory to the field to realize their potential. In this section we overview quantum network testbeds that have made this important jump. In line with the overall scope of this review, we focus on fiber-optic testbeds rather than free-space, yet we note the importance of satellite quantum communications for long-distance networking, particularly in the current pre-repeater era. Of multiple efforts globally~\cite{Sidhu2021},  the Micius mission has realized especially groundbreaking experimental demonstrations of satellite-to-ground QKD and entanglement distribution; we recommend a recent review \cite{Lu2022micius} for a summary of the progress so far.

When transferring a fiber-optic quantum communications experiment from the tabletop to a deployed network, the key challenges can be classified in two broad categories: (i) those related to the quantum channel, and (ii) those related to node synchronization. Examples of (i) include temperature variations, strain, and even air currents subjected to deployed fiber that a fiber spool in a well-controlled laboratory does not face, while time-stamping detection events and synchronizing coherent operations between distributed sites represent examples of (ii). A common strategy in quantum communication experiments considers loopbacks in which the send and receive connections of a deployed fiber are terminated in the same location; while this arrangement is sensitive to real-world channel impairments, it sidesteps the synchronization problem altogether, giving all ostensive nodes access to the same local clock for modulation and coincidence counting\footnote{Such loopbacks are distinct from bidirectional use of a single fiber, as in plug-and-play QKD, where Alice attenuates and reflects light back to Bob. In this case, Alice must synchronize her modulation modulation signal via photodetection~\cite{Stucki2002}.}. Given this significant simplification for loopbacks, we concentrate our review on testbeds with truly distributed sites, as validated by the measurement of entanglement or generation of a secret key in separate labs. For the sake of consistency, we report all distances from prior experiments in terms of deployed fiber lengths, excluding any extensions from additional spools. Note that many experiments referenced below also mention geographic distance---a number of particular importance in closing the locality loophole in Bell tests---which can differ significantly from the in-fiber number.

Remarkably, the first deployed quantum network appeared over two decades ago, realizing QKD between multiple nodes spread across BBN Technologies, Harvard University, and Boston University~\cite{elliott2002building, Elliott2005, Elliott2007}. Funded by DARPA, this QMAN ($\sim$20~km maximum distance) tested both weak-coherent-state and entanglement-based QKD, used the quantum key material for IPsec tunnels, and explored trusted nodes for extending distance without quantum repeaters. Bright wavelength-multiplexed optical pulses provided the needed synchronization and polarization tracking, while training frames were used for interferometer path-length stabilization. This testbed pioneered solutions to many of the engineering challenges in quantum networking and foreshadowed the appearance of a variety of deployed metro-scale QKD testbeds in cities throughout the world, including Vienna~\cite{Peev2009}; Hefei~\cite{Chen2009, Chen2010} and Wuhu~\cite{Wang2010} in China; Durban, South Africa~\cite{Mirza2010}; Geneva~\cite{Stucki2011}; Tokyo~\cite{Sasaki2011}; Cambridge, England~\cite{Dynes2019}; Madrid~\cite{Lopez2020}; and Chattanooga, Tennessee~\cite{Evans2021}. 

Despite the diversity in geography and applications of each of these testbeds, several trends have emerged. First, although entanglement-based QKD~\cite{Elliott2007, Peev2009, Sasaki2011, Evans2021} and CV-QKD~\cite{Peev2009, Lopez2020} have been considered in several of them, decoy-state BB84 with phase or polarization encoding has proven by far the most common protocol implemented, a testament to both its simplicity and performance. Second, nearly all testbeds have leveraged some form of wavelength-multiplexed classical pulses (CWDM or DWDM) for clock recovery and timing synchronization. And finally, all have supported fiber link distances on the order of $\sim$10--100~km, with trusted nodes enlisted to extend to distances beyond those of elementary links. Pushing the trusted node paradigm even further has spawned true QWANs that connect multiple QMANs, first joining the Hefei and Wuhu QKD networks~\cite{Wang2014} and more recently uniting QMANs in Beijing, Jinan, Hefei, and Shanghai~\cite{Chen2021}. In the latter case, nodes as far as 2000~km apart in fiber (4600~km including the satellite extension) can establish secret keys---far and away the longest distance of any deployed network.

Indeed, through the continued extension of trusted nodes, QKD on a global scale is possible with current technology. Yet trusted nodes are limited; not only does the paradigm rely on security assumptions that become increasingly less tenable for nodes physically and administratively distant from users, but it fundamentally  applies only to quantum networking use cases that do not require end-to-end entanglement. Irrespective of the quantum resources consumed, the output of QKD---namely, a shared secret key---is entirely classical and therefore  indefinitely storable in a conventional memory at a trusted node for future binary manipulation. On the other hand, quantum states cannot be stored classically, demanding quantum repeaters to establish entanglement over QWAN distances. Thus while quantum repeater research continues at an aggressive pace, near-term entanglement-based quantum networks are currently focusing on direct photonic transmission between nodes.

Entanglement distribution in deployed fiber was demonstrated first in Geneva, with time-bin-entangled photons distributed from a single source to two locations through 8.1~km and 9.3~km fiber channels~\cite{Tittel1998}. In both this and a subsequent nonlocality test~\cite{Salart2008}, coincidences were obtained via classical transmission of the detector output signals back to the source, thus supporting spatially distributed detection without explicit time synchronization. Other early work distributing polarization entanglement in deployed fiber appeared in Vienna~\cite{Poppe2004, Treiber2009}; %leveraging optical pulses for synchronization; 
enlisted for entanglement-based QKD, this system complemented the coherent-state systems also supported in the larger QKD testbed~\cite{Peev2009}. Recent work has pushed polarization entanglement distribution to fiber separations of 96~km in submarine~\cite{Wengerowsky2019} and 248~km in terrestrial~\cite{Neumann2022} links, while time-bin entanglement has recently been measured over 50~km in deployed fiber~\cite{Pelet2023}.

While most testbeds are focused on the distribution of DV entanglement, CV networks could prove valuable, especially in the context of quantum sensor networks.  In particular, distributed multi-mode squeezing could be a useful network resource. To this end, part of the Oak Ridge quantum network (discussed below) has distributed two-mode squeezing of over 1.5 km of deployed optical fiber in the C-band~\cite{Chapman23a}. 

While photonic quantum  teleportation can be enabled with DV or CV resources, it has not yet attained the same distances as elementary entanglement distribution. Time-bin qubits have nonetheless been teleported 0.8~km~\cite{Landry2007}, 17.3~km~\cite{Valivarthi2016}, and 30.4~km~\cite{Sun2016} in terms of total deployed fiber distance from Alice to Bob. On the matter qubit front, NV centers separated by 1.7~km of deployed fiber were entangled in one of the seminal loophole-free Bell tests~\cite{Hensen2015}, with a recent extension to 25~km enabled by quantum frequency conversion~\cite{Stolk2024}; atomic ensembles have been entangled over 17~km of deployed fiber~\cite{XiYu2022}, as well as individual atoms separated by 700~m~\cite{Zhang2022, vanLeent2022}; ion-ion~\cite{Krutyanskiy2023} and ion-photon~\cite{Kucera2024} entanglement has been realized over 520~m and 14.4~km, respectively; a photon entangled with a rare-earth quantum memory has been measured after 44~km~\cite{Rakonjac2023}; and two SiV spin qubits connected by a 35~km loopback (though located in separate rooms) have been entangled as well~\cite{Knaut2024}.

These deployed examples of entanglement distribution highlight the practical potential of quantum networks on metro-scales (and beyond), even in the near-term quantum-repeater-less regime. Yet although some of the above examples include three functional nodes, the third node is not itself a communicating party, but rather supplies entanglement or performs a joint measurement for two other users. Consequently, from a networking perspective, these experiments describe two-node communication \emph{links} and cannot be described as true multinode \emph{networks} characterized by at least three supported users. 

Testbeds throughout the world are pushing toward the multinode regime. For the purposes of this review, we focus on two in particular---at Delft University of Technology~\cite{Pompili2021, Pompili2022, Stolk2024} and Oak Ridge National Laboratory (ORNL)~\cite{Alshowkan2021, Alshowkan2022a, Alshowkan2022b, Alnas2022, Alshowkan2024}---that have pioneered distributed multinode entanglement\footnote{As three-node entanglement-based  demonstrations are appearing in the literature at a rapid pace~\cite{Rahmouni2024, Liu2024, Laneve2024}, our intent with the Delft-ORNL focus is not to exclude other leading testbeds, but rather highlight important engineering considerations relevant to hybrid classical-quantum networks as a whole.}.
The Delft testbed is based on NV center qubits; the Oak Ridge testbed is based on photonic qubits only, but is directly compatible with the lightwave communications infrastructure: operating in the C-band, using WSSs, and including a 1.2~km link separation in standard SMF. Both rely on quantum stacks analogous to TCP/IP and have explored the interplay between state quality and state quantity, via either a fidelity/latency tradeoff~\cite{Pompili2021, Pompili2022} or distillable entanglement~\cite{Alshowkan2021, Alshowkan2022a, Alshowkan2022b, Alnas2022}. Such network metrics highlight the desire to move beyond QKD to more general network applications in sensing and computing; indeed, demonstrations of remote state preparation (RSP) on both networks~\cite{Alshowkan2021, Pompili2022} point specifically to applications in blind quantum computing~\cite{Broadbent2009, Barz2012, Fitzsimons2017}.

As an alternative to the dedicated optical sync channels characteristic of most QKD testbeds, the ORNL QLAN employed Global Position System (GPS) clocks \cite{Alshowkan2021} and then later upgraded to White Rabbit \cite{Alshowkan2022b} for timing synchronization. Developed at CERN and incorporated into the Precision Time Protocol standard~\cite{IEEE2020}, White Rabbit is an open source solution for clock distribution over optical fiber, reaching remarkably low jitters on the order of 10~ps or less~\cite{Lipinski2011, Rizzi2018}. Commercially available, White Rabbit has been demonstrated over distances up to 950~km (with amplification)~\cite{Dierikx2016},  successfully daisy-chained up to 19 nodes~\cite{Gutierrez2021}, and characterized in detail for coexistence with C-band quantum signals~\cite{Burenkov2023}. White Rabbit therefore appears a particularly scalable approach for quantum network synchronization in lieu of traditional custom optical setups, finding adoption in other emerging quantum testbeds~\cite{Du2024, Rahmouni2024}. Yet we do acknowledge alternatives based on custom electronics~\cite{Valivarthi2022} and photon-pair-based coincidence tracking %and compensation of low-noise reference clocks
~\cite{Hughes2005, Ho2009, Shi2020, Pelet2023} receiving interest in this context as well.

\begin{figure}[t]
	\includegraphics[width=\textwidth]{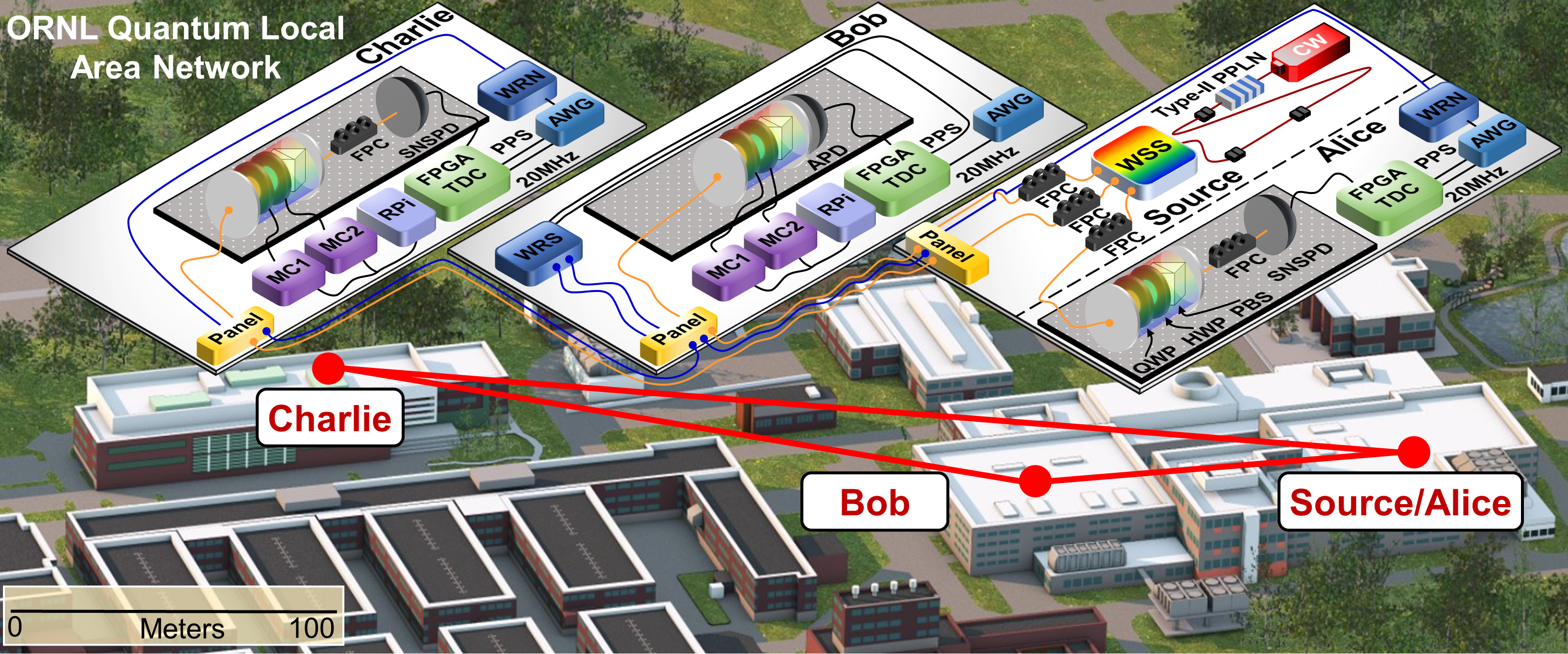}
    \centering
    %%	\captionsetup{justification=raggedright, singlelinecheck=false }
	\caption{Quantum network testbed at Oak Ridge National Laboratory. APD: avalanche photodiode; AWG: arbitrary waveform generator (for 10~MHz to 20~MHz clock doubling); CW: continuous-wave laser; FPC: fiber polarization controller; FPGA: field-programmable gate array; HWP: half-wave plate; MC: motion controller; PBS: polarizing beamsplitter; PPLN: periodically poled lithium niobate; PPS: pulse per second; QWP: quarter-wave plate; RPi: Raspberry Pi microprocessor board; SNSPD: superconducting nanowire single-photon detector; TDC: time-to-digital converter; WRN: White Rabbit node; WRS: White Rabbit switch; WSS: wavelength-selective switch.} 
	\label{fig:J3}
\end{figure}

Figure~\ref{fig:J3} depicts the components in the three primary quantum nodes of the ORNL QLAN. Polarization-frequency hyperentangled photons generated at Alice are spectrally partitioned by a WSS into three fiber outputs; one is kept at Alice and the other two sent to Bob and Charlie, each of whom possesses a polarization analyzer and detector for single-qubit measurements. Parallel fiber strands carry WR timing signals between a WR switch (WRS) at Bob and the two WR nodes (WRNs) at Alice and Charlie, which lock FPGA-based time taggers for coincidence detection. Finally, a computer at Alice (omitted for simplicity) communicates over ethernet with a Raspberry Pi (RPi) at Bob and Charlie for instrument control and data collection. Even in this relatively simple network, the integration of quantum resources (embodied in the generation, distribution, and measurement of entangled photons) with classical resources (represented by WR synchronization, time-tagging electronics, and parallel classical communications) reinforces a main theme of this review: quantum networking is inherently \emph{hybrid}, requiring both quantum and classical network technologies acting in concert for successful implementation.

There remain many directions for improvement on existing networks like that in Fig.~\ref{fig:J3}. In addition to expanding distance after the advent of quantum repeaters, an important target is the integration of matter qubits with a telecom fiber infrastructure.
Central to such a vision are quantum transducers, devices that coherently convert quantum information from one physical system to another~\cite{Lauk2020, Awschalom2021, Han2021}. Ultimately, it is desirable that such transducers would function fault tolerantly. We are excited by the sustained improvement in quantum transducers spanning the particularly difficult microwave-to-optical energy regimes, leveraging physical mechanisms ranging from optomechanical~\cite{Mirhosseini2020} and electro-optic resonators~\cite{Sahu2022} to neutral atoms~\cite{Kumar2023} and ions~\cite{Xie2025}. 
To the extent quantum memories supporting direct telecom-band photon storage are available---also an active  research front~\cite{Saglamyurek2015, Lei2023}---transduction between widely different energies could potentially be averted, at least in the core of the optical network.

From an application perspective, a multinode network demonstration of quantum-enhanced capabilities \emph{other} than QKD---such as entanglement-enhanced sensing~\cite{xia2020demonstration,xia2023entanglement,guo2020distributed,liu2021distributed}, entanglement-enhanced telescopy~\cite{gottesman2012longer,
brown2023interferometric}, distributed quantum computing~\cite{Cirac1999, Cacciapuoti2020}, or blind quantum quantum cloud computing~\cite{Broadbent2009}---would mark a critical milestone in the field, potentially heralding an era of quantum networks that are valued primarily for their practical value than their research interest.

Table~\ref{tab:entDist} concludes this subsection with a summary of key entanglement distribution experiments mentioned above.  %shows the geographic location and salient features of these demonstrations, highlighting aspects such as the number of nodes, maximum fiber distance, and implemented protocols. 
In keeping with this review's deployed focus, we include only experiments in which the entangled particles are measured nonlocally (i.e., without looping back into the same laboratory) and %---as mentioned earlier in Sec.~\ref{sec:qNetDemo}---
define the maximum distance as the length of the fiber lightpath between the farthest entangled nodes.
In our view, this distance definition is most relevant to hybrid classical-quantum networking, where we typically envision building a quantum network on top of existing optical fiber infrastructure (nodes and fiber paths). %have already been established.
In such a context, %the fiber lightpath itself most directly impacts design considerations, for 
the shortest ``straight-line'' distance is not generally available, and additional fiber spools---often inserted to add delay or simulate longer links---unnecessarily reduce performance in a application-focused network.
Entanglement is classified by particle type: flying photonic qubits or stationary memory qubits  (e.g., color centers, atoms, or ions). The most advanced protocol implemented in each reference comprises the fifth column, from baseline entanglement distribution to entanglement swapping or an application like QKD. %For organizational purposes, we order the results by date of publication, which of course does not always reflect the timeline of the demonstrations themselves.

Analogous to the Internet Protocol itself, we have curated Table~\ref{tab:entDist} as a ``best effort'' service; %with the goal of offering a broad flavor of the field's development; 
no guarantee of exhaustivity is made. Nevertheless, perhaps the most striking feature is the explosion of results in the last three years. Although covering a 26-year span (1998--2024), more than half of all references in Table~\ref{tab:entDist} have appeared since 2022, providing yet another indication of the growing interest in quantum networking globally.

\begin{table}[!tb]
\centering
\footnotesize
\caption{Summary of entanglement distribution experiments in deployed fiber.}
\label{tab:entDist}
\begin{tabular}{|c|c|c|c|c|c|}
\hline
         & \#         & Max. fiber        &              &          & Year \\
Location & nodes & distance [km] & Entanglement & Experiment & published \\\hline
Geneva & 3 & 17.4 & photon/photon & Bell test & 1998~\cite{Tittel1998} \\ % Oct
Vienna & 2 & 1.45 & photon/photon & QKD & 2004~\cite{Poppe2004} \\ % Aug
Geneva & 2 & 0.8 & photon/photon & teleportation & 2007~\cite{Landry2007} \\ % Feb
Geneva & 3 & 30.9 & photon/photon & Bell test & 2008~\cite{Salart2008} \\ % Aug
Vienna & 2 & 16 & photon/photon & QKD & 2009~\cite{Treiber2009} \\ % Apr
Delft & 3 & 1.7 & memory/memory & Bell test & 2015~\cite{Hensen2015} \\ % Oct
Hefei & 3 & 30.4 & photon/photon & teleportation & 2016~\cite{Sun2016} \\ % Sep 19
Calgary & 3 & 17.3 & photon/photon & teleportation & 2016~\cite{Valivarthi2016} \\ % Sep 19
Hefei & 3 & 25.8 & photon/photon & swapping & 2017~\cite{Sun2017} \\ % Oct
Malta--Sicily & 2 & 96 & photon/photon & distribution & 2019~\cite{Wengerowsky2019} \\ % Apr
Delft & 3 & 0.03 & memory/memory & RSP & 2021~\cite{Pompili2021} \\ % Apr
Oak Ridge & 3 & 1.5 & photon/photon & RSP & 2021~\cite{Alshowkan2021} \\ % Oct
Munich & 3 & 0.7 & memory/memory & swapping & 2022~\cite{vanLeent2022}\\ % Jul 6
Munich & 2 & 0.7 & memory/memory & QKD & 2022~\cite{Zhang2022}\\ % Jul 27
Hefei & 2 & 17 & memory/memory & distribution & 2022~\cite{XiYu2022} \\ % Jul 28
Austria--Slovakia & 3 & 248 & photon/photon & distribution & 2022~\cite{Neumann2022} \\ % Oct
Innsbruck & 2 & 0.52 & memory/memory & swapping & 2023~\cite{Krutyanskiy2023} \\ % Feb
Nice%--Sophia Antipolis
& 3 & 50 & photon/photon & QKD & 2023~\cite{Pelet2023} \\ % Oct
Barcelona & 2 & 44  & photon/memory & distribution & 2023~\cite{Rakonjac2023} \\ % Dec
Oak Ridge & 3 & 1.5 & photon/photon & digital signatures & 2024~\cite{Chapman2024dig} \\ % Feb
DC Metro & 3 & 0.25 & photon/photon & distribution & 2024~\cite{Rahmouni2024} \\ % Mar
Hefei & 3 & 21.6 & memory/memory & swapping & 2024~\cite{Liu2024} \\ % May
Oak Ridge & 6 & 2.1 & photon/photon & distribution & 2024~\cite{Alshowkan2024} \\ % Jul
Saarbr\"{u}ken & 2 & 14 & photon/memory & teleportation & 2024~\cite{Kucera2024} \\ % Sep
Delft & 3 & 25 & memory/memory & swapping & 2024~\cite{Stolk2024} \\ % Oct
Chattanooga & 3 & 6.7 & photon/photon & distribution & 2024~\cite{Chapman2024} \\ % Dec
Tempe & 2 & 1.6 & photon/photon & process tomography & 2025~\cite{Rahman2025} \\ 
\hline
\end{tabular}
\end{table}

\subsection{Coming full circle: quantum technology for classical communications}

Recurring themes throughout this review have emphasized the value of leveraging existing technology from classical lightwave communications for quantum communications---unsurprising given the maturity of optical networks and the importance of integrating quantum systems into the existing fiber-optic infrastructure for practical deployment. Nevertheless, continued advances in quantum information processing systems lead us to predict a future filled with technology transfer in the \emph{opposite} direction as well, as tools motivated by and developed for quantum networking find new applications in classical contexts, auspiciously benefiting the lightwave communications field that originally supported it. In this picture of the future, acute needs of quantum networks inspire solutions which in turn address complementary needs in classical networking that have hitherto lacked sufficient economic drivers.

There are several areas where the development of new quantum networking technology could be of benefit to conventional networks.  In particular, lower-loss and lower-noise optical components could enable reduced DSP to correct for transmission impairments, leading to improved energy efficiency. Low-loss optical circuitry represents one such example immediately applicable to classical use. As highlighted in Sec.~\ref{sec:quantumIncompatible}, conventional optical amplification is unavailable to quantum signals~\cite{Wootters1982,dieks1982communication}. Consequently, loss cannot be compensated via EDFAs in quantum communications, fundamentally limiting the distance over which quantum information can be transmitted. In this milieu, multi-dB insertion losses---e.g., from electro-optic modulators, switches, and pulse shapers---that are perhaps acceptable inconveniences in classical communications suddenly become critical roadblocks in the quantum regime. On the quantum computing front, companies like PsiQuantum~\cite{Alexander2024} are already driving the development of low-loss photonic integrated circuits, and we predict quantum networking will increasingly motivate research in low-loss communications components, both integrated and discrete. Such technology would then readily translate to classical communication links, enabling energy and cost savings by removing amplification or lowering launch powers.

The excess loss for optical components as well as the actual fiber transmission loss in telecommunications systems is generally budgeted to allow for sufficient receive power to enable low-error operation even with variations in the specifications of optical systems.  Cost has always been a major driver in the telecommunications industry, and as a result, even if lower-loss versions of optical components are available, they have traditionally been more expensive ``premium grade'' parts and their use is minimized.  However, as data demands have required better spectral efficiency, on-off keyed modulation formats have given way to coherent modulation in amplitude and phase.  If low-loss optical components were the norm, they could perhaps be utilized to reduce the number of optical amplifiers in a given deployment, which opens the door for reduced signal launch power, and potentially improved energy efficiency.  

Among the various low-loss optical components, one particularly promising technology for both classical and quantum communication is the hollow core nested antiresonant nodeless fiber (HC-NANF)~\cite{poletti2014nested}. HC-NANFs offer several significant advantages over conventional glass-core fibers, including reduced optical nonlinearity, ultralow latency, ultrawide bandwidths, and the potential of lower transmission loss. Notably, the first hollow core fiber (HCF) to achieve loss levels (0.174 dB/km) comparable to conventional telecom fibers in the C-band was reported in~\cite{jasion20220}. The reduced nonlinearity of HC-NANFs results in lower crosstalk between classical and quantum channels in quantum/classical coexistence networks. This was demonstrated in a recent experiment where DV-QKD was successfully conducted alongside 1.6 Tbps classical traffic over a HCF link~\cite{alia2022dv}. Although HC-NANF was not specifically designed for quantum networking, its potential in future quantum Internet applications may justify the cost of replacing existing optical fibers.

Given the central role of entanglement in quantum networking, a particularly promising advancement in classical communication enabled by quantum technologies is entanglement assisted communication (EACOMM). By utilizing preshared entanglement, EACOMM can significantly enhance classical communication rates~\cite{bennett2002entanglement}, as exemplified by superdense coding~\cite{bennett1992communication,mattle1996dense,barreiro2008beating,williams2017superdense}. However, such schemes still operate within the limits imposed by the Holevo--Schumacher--Westmoreland (HSW) bound ~\cite{hausladen1996classical,schumacher1997sending,holevo2002capacity}, which defines the ultimate capacity for classical information transmission over quantum channels. More recently, by harnessing phase correlations between entangled bosonic modes, EACOMM surpassing the HSW bound has been demonstrated experimentally~\cite{hao2021entanglement}, marking a potentially transformative shift in communication theory. Although the generation and distribution of entanglement between remote parties remain challenging and resource-intensive, one can envision future hybrid classical-quantum networks where entanglement is distributed during off-peak hours, allowing for boosted classical communication efficiency during high-traffic periods.

On the architecture side, quantum information's incompatibility with traditional OE conversion could foreshadow a renaissance in all-optical signal processing and packet switching, a possibility speculated in Secs.~\ref{sec:allOptical} and \ref{sec:quantumMultiuser}. Efficient error-correcting quantum memory along with better network synchronization might also enable advanced concepts of all-optical classical networking. Given the more stringent constraints on noise and power levels in quantum applications, we anticipate that quantum all-optical solutions should generally transfer to the classical domain where they could increase bandwidth, lower latency, and reduce energy use. Indeed, illustrations of such quantum-to-classical translation have already emerged; we highlight one example in the context of the QFP introduced briefly in Sec.~\ref{sec:quantumMultiuser}. Based on electro-optic phase modulators and Fourier-transform pulse shapers, the QFP enables the scalable synthesis of arbitrary quantum gates in frequency-bin encoding~\cite{Lu2019c}---a particularly appropriate DoF for stable and parallel transmission in SMF~\cite{Kues2019, Lu2019c, Lu2023a, Lu2023c}.

Although inspired by and focused on quantum information processing in the majority of its experimental demonstrations, the QFP at bottom is a mode transformer: it coherently mixes and interferes discrete frequency bins in a controllable fashion. This mode transformation is optically linear and entirely independent of the input's statistical properties, making the QFP equally accepting of classical states as quantum ones. Recognizing the opportunities presented by the QFP's flexibility, previous research has explored classical all-optical processing both theoretically~\cite{Lukens2020a} and experimentally~\cite{Lu2020a} for the specific operations of frequency broadcasting (where data at one input frequency is copied to $N$ outputs) and cyclic hops (where data at multiple input frequencies are transferred to distinct outputs). Reconfigurable, background-free, and based on the theoretical guarantees of the QFP, this all-optical approach can in principle support arbitrary wavelength conversion operations in a single platform, highlighting by example the value of considering quantum and classical applications together as quantum network technology evolves; their intersection is ripe for new ideas and crosspollination that can advance both.

\section{Outlook}
\label{sec:outlook}

As the backbone of the modern Internet, fiber-optic communication networks have profoundly shaped human society and civilization. To date, most optical communication networks operate in \textit{classical} communication modes, where information carried by light is fundamentally distinguishable. The emerging field of quantum communication expands the realm of optical communication by leveraging the generation, transmission, and detection of quantum superposition states of photons. This development unlocks revolutionary applications such as distributed quantum computing and ultrasecure communications. In this context, the envisioned \textit{quantum Internet} promises a broader scope than today's classical Internet, supporting both quantum and classical communication modalities.

However, transitioning existing optical networks to support quantum communication presents immense challenges. A fundamental task of quantum communication—deterministically transmitting a quantum state through a lossy channel—cannot be achieved using classical optical solutions such as regenerators and amplifiers. The key obstacle stems from the quantum no-cloning theorem, which prohibits the perfect duplication of unknown quantum states. Although the concept of quantum repeaters has been proposed, and preliminary experimental progress has been made, practical quantum repeaters remain years away from realization. In the near term, quantum communication protocols that tolerate probabilistic transmissions, such as QKD, are the only feasible solutions. Given the niche market for these early applications, the most pragmatic commercialization approach involves integrating quantum technologies into existing fiber-optic networks. Such integration will extend beyond the physical layer, encompassing the entire network stack.

The pursuit of a global quantum Internet will drive the development of cutting-edge quantum technologies, including entangled photon sources, quantum memories, and quantum repeaters. Simultaneously, advancements in existing optical technologies---such as low-noise optical receivers, ultralow-loss optical storage, novel optical fibers, and related components---will push the boundaries of classical communication. These innovations may enable previously impractical modalities, such as all-optical networks, to become viable solutions. In the long term, the Internet of the future is likely to consist of heterogeneous communication networks designed to seamlessly support both classical and quantum communication, ensuring a robust, versatile, and transformative global communication infrastructure.

\textbf{Credit authorship contribution statement}

Joseph M. Lukens: Writing – review \& editing, Writing – original draft. Nicholas A. Peters: Writing – review \& editing, Writing –
original draft. Bing Qi: Writing – review \& editing, Writing – original draft. 

\textbf{Declaration of competing interest}

The authors declare that they have no known competing financial interests or personal relationships that could have appeared
to influence the work reported in this paper.

\textbf{Acknowledgments}

We are grateful to Li Qian for helpful discussions while preparing this review and to Muneer Alshowkan for providing the image in Fig.~\ref{fig:J3}. B.Q. acknowledges support from NYU-Shanghai start-up funds. J.M.L. recognizes support from the U.S. Department of Energy (ERKJ432,
DE-SC0024257) and Sandia National Laboratories (Laboratory Directed Research and Development Program). This work
was performed, in part, at Oak Ridge National Laboratory, operated by UT-Battelle for the U.S. Department of energy under contract no. DE-AC05-00OR22725. N.A.P. recognizes support from the U.S. Department of Energy, Office of Science, Advanced Scientific Computing Research (ERKJ381, ERKJ378).

\bibliographystyle{elsarticle-num}
\bibliography{main}

\end{document}